\documentclass[structabstract]{aa}

\usepackage{natbib}                     
\usepackage{graphicx}                   
\usepackage[varg]{txfonts}              
\usepackage{hyperref}             
\hypersetup{breaklinks=true,colorlinks=true,urlcolor=blue, linkcolor=blue,  citecolor=blue,pagecolor=red}

  \def \teff {$T_{\mathrm{eff}}$}
  
  \def \vtur {$V_{\mathrm{tur}}$}
  \def \logg {$\log g$}

\begin{document}

  \title{Benchmark stars, benchmark spectrographs}
  \subtitle{Detailed spectroscopic comparison of ESPRESSO, PEPSI, and HARPS data \\ for Gaia benchmark stars\thanks{Based on observations collected at the Paranal Observatory, ESO (Chile) with the ESPRESSO spectrograph at the VLT (ESO runs ID 0102.D-0185(A); 0103.D-0118(A); 0104.D-0362(A)).}}

  \author{V.~Adibekyan \inst{1,2} \and
          S.~G.~Sousa \inst{1} \and
          N.~C.~Santos \inst{1,2} \and
          P.~Figueira\inst{3,1} \and
          C.~Allende~Prieto \inst{4,5} \and
          E.~Delgado~Mena \inst{1} \and
          J.~I.~Gonz\'alez~Hern\'andez \inst{4,5} \and
          P.~de~Laverny\inst{6} \and
          A.~Recio--Blanco\inst{6} \and
          T.~L.~Campante \inst{1,2} \and
          M.~Tsantaki \inst{7} \and
          A.~A.~Hakobyan\inst{8} \and  
          M.~Oshagh \inst{4,5} \and
          J.~P.~Faria\inst{1,2} \and
          M.~Bergemann \inst{9} \and
          G.~Israelian \inst{4,5} \and
          T.~Boulet \inst{1,2}
         }

  \institute{
          Instituto de Astrof\'isica e Ci\^encias do Espa\c{c}o, Universidade do Porto, CAUP, Rua das Estrelas, 4150-762 Porto, Portugal \\
          \email{vadibekyan@astro.up.pt} \and
          Departamento de F\'{\i}sica e Astronomia, Faculdade de Ci\^encias, Universidade do Porto, Rua do Campo 
          Alegre, 4169-007 Porto, Portugal \and
          European Southern Observatory, Alonso de Córdova 3107, Vitacura, Región Metropolitana, Chile \and
          Instituto de Astrof\'{i}sica de Canarias, E-38205 La Laguna, Tenerife, Spain \and
          Departamento de Astrof\`{i}sica, Universidad de La Laguna, E-38206 La Laguna, Tenerife, Spain \and
          Université C\^{o}te d'Azur, Observatoire de la C\^{o}te d'Azur, CNRS, Laboratoire Lagrange, Bd de l'Observatoire, CS 34229, 06304 Nice cedex 4, France \and
          INAF -- Osservatorio Astrofisico di Arcetri, Largo E. Fermi 5, 50125, Firenze, Italy \and
          Center for Cosmology and Astrophysics, Alikhanian National Science Laboratory, 2 Alikhanian Brothers Str., 0036 Yerevan, Armenia \and
          Max Planck Institute for Astronomy, 69117, Heidelberg, Germany
          }

  \date{Received date / Accepted date }
  \abstract
  {Gaia benchmark stars are selected to be calibration stars for different spectroscopic surveys. Very high-quality and homogeneous spectroscopic data for these stars are therefore required. We collected ultrahigh-resolution ESPRESSO spectra for 30 of the 34 Gaia benchmark stars and made them public.}
  {We quantify the consistency of the results that are obtained with different high- (R $\sim$ 115\,000), and ultrahigh- (R $\sim$ 220\,000) resolution spectrographs. We also comprehensively studied the effect of using different spectral reduction products of ESPRESSO on the final spectroscopic results.}
  {We used ultrahigh- and high-resolution spectra obtained with the ESPRESSO, PEPSI, and HARPS spectrographs to measure spectral line characteristics (line depth; line width; and equivalent width, EW) and determined stellar parameters and abundances for a subset of 11 Gaia benchmark stars. We used the ARES code for automatic measurements of the spectral line parameters.}
  {Our measurements reveal that the same individual spectral lines measured from adjacent 2D (spectrum in the wavelength-order space) echelle orders of ESPRESSO spectra differ slightly in line depth and line width. When a long list of spectral lines is considered, the EW measurements based on the 2D and 1D (the final spectral product) ESPRESSO spectra agree very well. The EW spectral line measurements based on the ESPRESSO, PEPSI, and HARPS spectra also agree to within a few percent. However, we note that the lines appear deeper in the ESPRESSO spectra than in PEPSI and HARPS. The stellar parameters derived from each spectrograph by combining the several available spectra agree well overall.}
  {We conclude that the ESPRESSO, PEPSI, and HARPS spectrographs can deliver  spectroscopic results that are  sufficiently consistent for most of the science cases in stellar spectroscopy. However, we found small but important differences in the performance of the three spectrographs that can be crucial for specific science cases.}
  
  \keywords{Stars: fundamental parameters -- Techniques: spectroscopic -- Stars: abundances -- Instrumentation: spectrographs}


  \maketitle
   
\section{Introduction}                                  \label{sec:intro}

Stellar spectra contain a wealth of information about the formation, evolution, and current characteristics of stars. The amount and robustness of the information that can be extracted from spectra  depends on their quality: the higher the signal-to-noise ratio (S/N)  and spectral resolution (R), the more detailed and precise the extracted information. In addition to these properties, the homogeneity of the data and analysis method becomes crucial for comparative studies among several stars. These obvious factors (high quality and homogeneity) are the base for the success of small and large surveys. Representative examples of small programs are the solar-twin observations made with the High Accuracy Radial velocity Planet Searcher (HARPS)  by \citet{Nissen-15}, the HARPS Guaranteed Time Observations (GTO) survey for exoplanet and galactic stellar populations studies \citep{Adibekyan-13, Adibekyan-12_harps}, and the AMBRE\footnote{The Archeologie avec Matisse Bas\'{e}e sur les aRchives de l'ESO.} project \citep{deLaverny-12}. The large successful spectroscopic surveys include the Gaia-ESO \citep{Gilmore-12} and the Apache Point Observatory Galactic Evolution Experiment \citep[APOGEE -][]{Majewski-17} surveys, to mention a few.

Several additional very large ground-based spectroscopic surveys of Milky Way stars are ongoing, with the common goal of understanding the formation and structure of our Galaxy (e.g., GALAH\footnote{The GALactic Archaeology with HERMES.} \citep{DeSilva-15}, LAMOST\footnote{The Large Sky Area Multi-Object Fibre Spectroscopic Telescope.} \citep{Deng-12}, SDSS\footnote{The Sloan Digital Sky Survey.} \citep{Yanny-09}, and RAVE\footnote{The Radial Velocity Experiment.} \citep{Steinmetz-06}). The Gaia space mission \citep{Perryman-01} beautifully complements this list of ground-based surveys. Depending on the specifications of the science goals and instrument configuration, these surveys use different methods and techniques to provide their final scientific products (e.g., stellar parameters and chemical abundances). It is therefore not straightforward to combine scientific outputs from different surveys without common standard stars for calibration. 

A massive effort has been made by the Gaia-ESO scientific community to define a list of 34 FGK stars (including the Sun) and M giants that can be used as benchmarks \citep[][]{Heiter-15}. Most of these stars, called Gaia benchmark stars, have interferometric observations that allow fundamental, that is, less model-dependent and spectrum-independent,  methods to be used to determine their effective temperature and surface gravity. Five metal-poor candidates were later added to the list \citep{Hawkins-16}. However, these new targets do not have interferometric observations, like the initial Gaia benchmarks do. Because the spectra of these stars were obtained from different high-resolution (HR) spectrographs, they were first convolved and resampled to the lowest resolution  (first to R$\sim$70\,000 and later convolved to a common resolution of R$\sim$47\,000) of the spectrographs and to cover the visible range from 480 to 680 nm \citep{Blanco-Cuaresma-14}. The spectra from this  homogenized library were used for further spectroscopic analyses of the stars with different methods  by different groups \citep[see, e.g.,][]{Jofre-14, Jofre-15}. We refer to \citet{Jofre-17} for further details and references.

However, when very high precision in stellar parameters and chemical abundances are the aim, then even a higher resolution and  intrinsic homogeneity of the spectra might be required. An important work toward creating a truly homogeneous ultrahigh-resolution (UHR) library of Gaia benchmark and other standard stars has recently been performed by \citet{Strassmeier-18}. The authors observed  48 bright AFGKM stars, including 23 of the Gaia benchmark stars, with the Potsdam Echelle Polarimetric and Spectroscopic Instrument (PEPSI) spectrograph installed at the Large Binocular Telescope (LBT) in Arizona \citep{Strassmeier-15AN}. These spectra have an average R$\sim$220\,000, a wavelength coverage of  383--912 nm, and a S/N  in the combined spectra that ranges from 70 to 6\,000, depending on the star and wavelength region. In addition to achieving a very precise determination of global stellar parameters, the authors demonstrated the high potential of using these UHR spectra by detecting the rare-earth element dysprosium and other difficult-to-measure heavy elements, such as uranium and thorium.  The authors also measured  $^{12}$C/$^{13}$C isotope ratios, which have very important implications for different fields of stellar astrophysics and planet engulfment theories \citep[e.g.,][]{Privitera-16}, and even for the search for solar siblings \citep{Adibekyan-18}.

As mentioned above, only 23 of the 34 Gaia benchmark stars have been observed with PEPSI. This is because the majority of these benchmark stars are located in the southern hemisphere. The Echelle SPectrograph for Rocky Exoplanets and Stable Spectroscopic Observations (ESPRESSO) spectrograph \citep{Pepe-10, Gonzalez-Hernandez-18}, installed at the Very Large Telescope (VLT), presented us with the unique opportunity of observing the southern benchmark stars at a resolution higher than 220\,000. Our team seized this opportunity and observed 30 of the Gaia benchmark stars in a uniform fashion. The remaining four targets are not observable with ESPRESSO but have previously been observed with PEPSI. The science-ready 1D combined ESPRESSO spectra of the 30 Gaia benchmark stars are made public and can be downloaded from our dedicated webpage\footnote{\url{http://www.astro.up.pt/~vadibekyan/benchmark-espresso.htm}}. 

The PEPSI and ESPRESSO instruments are new, have different specifications (because their scientific goals are different), and are unique (spectral coverage and resolution), thus  it is important to verify whether these two benchmark spectrographs give consistent results. It has been shown that when stellar parameters are compared that were derived from different spectra, obtained  with different instruments (i.e., spectral resolution and wavelength coverage), and under different observing and meteorological conditions, the choice of the instrument plays a major effect \citep[see, e.g.,][]{Bedell-14, Bensby-14}. Although the resulting differences are acceptably small for most of the science cases \citep{Sousa-18}, they can be very important when very fine differences in chemical abundances between binary stars  \citep[e.g.,][]{Saffe-16, Adibekyan-16, Ramirez-19}, abundance differences between stars with and without planets \citep[e.g.,][]{Adibekyan-12_kepler, Adibekyan-15, Gonzalez-Hernandez-10, Gonzalez-Hernandez-13, Delgado-Mena-18}, or the composition of different galactic populations are studied \citep[][]{Nissen-10, Adibekyan-11, Bergemann-14, Spina-16, Nissen-20}.

The main aim of this work is to make a detailed  comparison of the spectral data products of ESPRESSO and PEPSI. We did not directly compare the spectra, but the properties of spectral lines (line depths; widths; and equivalent widths, EW). This is a very important test for the scientific community to know at which level these UHR spectrographs are compatible. In addition, we compared the spectroscopic results obtained with these two spectrographs with those taken with the HARPS spectrograph. HARPS has a spectral resolution of about 115\,000 (nearly half of the resolution of PEPSI and ESPRESSO) and has been and contiues to be one of the most frequently used ESO spectrographs since its commissioning in 2002 \citep{Mayor-03}.

The outline of this manuscript is as follows: in Sect.~\ref{sec:sample} we briefly describe the sample. At the beginning of  Sect.~\ref{sec:espresso} we briefly described the ESPRESSO spectrograph. Then, in Sects \ref{sec:multi_order}, \ref{sec:1d_vs_2d}, \ref{sec:sun_params}, \ref{sec:slices}, and \ref{sec:1d_1d} we compare the spectral line properties  in detail, based on  measurements from the ESPRESSO spectra taken for the Sun in different formats. The comparison of the  spectroscopic results based on the ESPRESSO, PEPSI, and HARPS spectra for the becnhmark stars is presented in Sect.~\ref{sec:spectrographs}. A summary of the results is provided in Sect.~\ref{sec:summary}.

\section{Sample}                                    \label{sec:sample}

\begin{figure}
\begin{center}
\begin{tabular}{c}
\includegraphics[angle=0,width=0.9\linewidth]{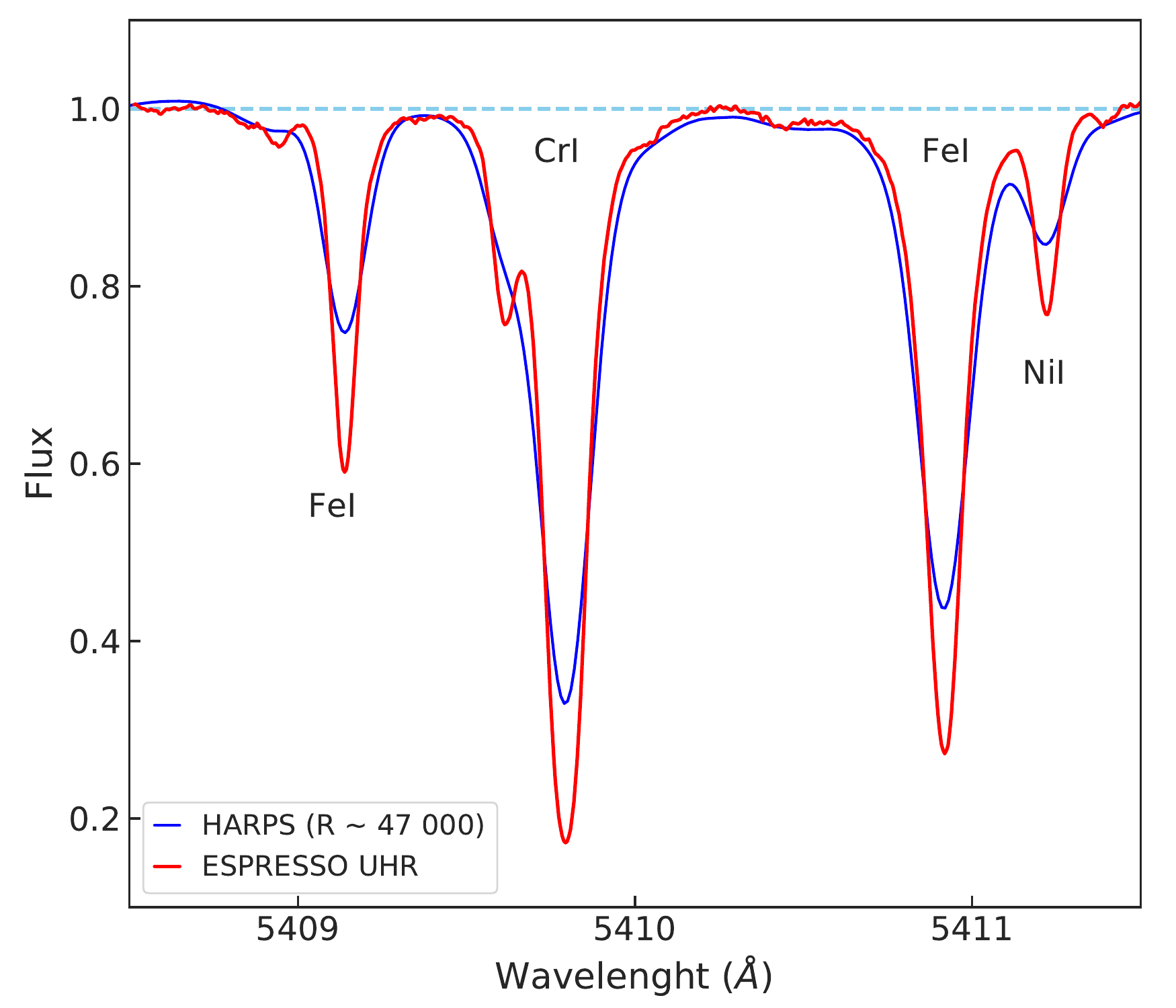}
\end{tabular}\end{center}
\vspace{-0.5cm}
\caption{ Comparison of the ESPRESSO UHR (R $\sim$ 220\,000) and HARPS spectra convolved to R=47\,000 of $\tau$ Cet. Only a small spectral region containing two commonly used FeI, CrI, and NiI lines is shown.}
\label{fig-taucet_espresso_harps}
\end{figure}

\begin{figure}
\begin{center}
\begin{tabular}{c}
\includegraphics[angle=0,width=0.95\linewidth]{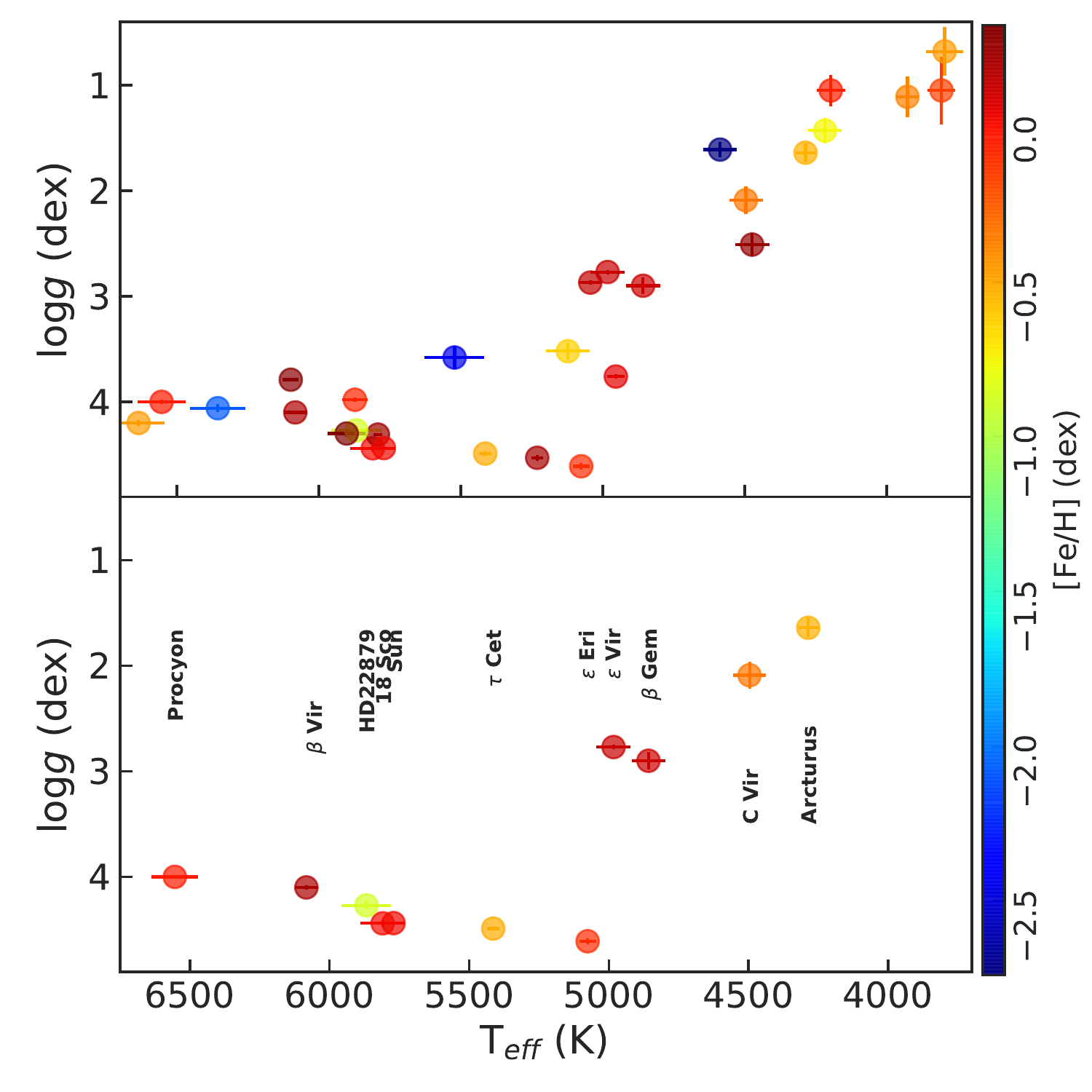}
\end{tabular}\end{center}
\vspace{-0.7cm}
\caption{Spectroscopic HR diagram for the 30 Gaia benchmark stars observed with ESPRESSO (top panel) and for the subsample of stars that were also observed with PEPSI and HARPS (bottom panel). The symbols are color-coded according to the stellar metallicity.}
\label{fig-hr-diagram}
\end{figure}

During ESO periods P102-P104 we observed 30 (out of 34) Gaia benchmark stars with the ESPRESSO spectrograph in UHR mode. In this mode the spectral resolution is four times higher than the nominal (minimum) resolution of the spectra collected in \citet{Blanco-Cuaresma-14}. Fig.~\ref{fig-taucet_espresso_harps} illustrates the effect of this difference in spectral resolution for $\tau$ Cet, the slowest rotating (v$\sin i <$ 1 km/s) star in our sample\footnote{We note that except this figure, we always used HARPS spectra in their original R $\sim$ 115\,000 resolution.}. The distribution of these stars in the Hertzsprung--Russell (HR) diagram is presented in Fig.~\ref{fig-hr-diagram}, and their stellar characteristics together with the S/N of the ESPRESSO spectra are presented in Table~\ref{tab:all_parameters}. Twenty of these 30 stars also have UHR PEPSI observations, and for 18 of them, HR spectra were obtained with HARPS.

From the 18 stars with ESPRESSO, PEPSI, and HARPS observations we excluded 3 stars (HD\,84937, HD\,140283, and HD\,122563) with [Fe/H] $<$ -2 dex, 2 stars ($\alpha$ Cet and $\alpha$ Tau) with \teff \ $<$ 4000 K, and 2 stars ($\eta$ Boo and HD\,49933) with v$\sin i$ $>$ 10 km/s. The reason for this was that we wished to avoid spectra with very weak (metal-poor stars), blended (cool stars), and very broad (fast rotators) spectral lines, which are difficult to measure with automatic spectral line fitting tools. The resulting sample of 11 stars is shown in the bottom panel of Fig.~\ref{fig-hr-diagram}, and their characteristics are presented in Table~\ref{tab:parameters}. The S/N at $\sim$5800 \AA{} for the HARPS and ESPRESSO combined spectra was calculated from the flux (photon counts) and for the PEPSI spectra, the S/N was taken from \citet{Strassmeier-18, Strassmeier-18_sun}. These values are presented in Table~\ref{tab:parameters} as well.

We used a list of about 460 spectral lines belonging to 28 different chemical species from carbon to neodymium \citep{Adibekyan-12_1111, Adibekyan-15a, Sousa-08, Delgado-Mena-17}. The spectral line properties were measured with the automatic ARES v2 code\footnote{The last version of the ARES code (ARES v2) can be downloaded \url{http://www.astro.up.pt/~sousasag/ares}} \citep{Sousa-15}. ARES performs a local continuum normalization around each spectral line and fits Gaussian profiles to measure the line properties.

\begin{table*}[t!]
\caption{\label{tab:parameters} Main properties of the benchmark stars with ESPRESSO, PEPSI, and HARPS observations. The stellar parameters are from \citet{Jofre-15}.}
\centering
\small
\begin{tabular}{llllllll}
\hline\hline
star & S/N$_{ESPRESSO}$ &   S/N$_{PEPSI}$ & S/N$_{HARPS}$ & \teff \ (K) & \logg \ (dex) & [Fe/H] (dex) & v$\sin i$  (km/s) \\
\hline
18 Sco  & 570 & 1040 & 400 &   5810$\pm$80  &  4.44$\pm$0.03  &  0.01$\pm$0.03  &  2.2$\pm$1.2\\ 
Arcturus  & 570 & 3000 & 400 &  4286$\pm$35  &  1.64$\pm$0.09  &  -0.53$\pm$0.08  &  3.8$\pm$1.0\\ 
Procyon  & 380 & 1040 & 1470 &  6554$\pm$84  &  4.00$\pm$0.02  &  -0.04$\pm$0.08  &  2.8$\pm$0.6\\ 
$\beta$ Vir  & 690 & 330 & 1230 &  6083$\pm$41  &  4.10$\pm$0.02  &  0.21$\pm$0.07  &  2.0$\pm$0.6\\
$\beta$ Gem  & 520 & 1850 & 280 &  4858$\pm$60  &       2.9$\pm$0.08  &    0.12$\pm$0.16  &       2.0$\pm$1.0 \\
$\epsilon$ Eri  & 1510 & 1800 & 1330 &  5076$\pm$30  &  4.61$\pm$0.03  &  -0.10$\pm$0.06  &  2.4$\pm$0.2\\ 
$\epsilon$ Vir  & 570 & 450 & 580 &  4983$\pm$61  &  2.77$\pm$0.02  &  0.13$\pm$0.16  &  2.0$\pm$1.0\\ 
HD 107328  & 600 & 1110 & 370 &  4496$\pm$59  &  2.09$\pm$0.13  &  -0.34$\pm$0.16  &  1.9$\pm$1.2\\ 
HD22879  & 600 & 540 & 1270 &  5868$\pm$89  &  4.27$\pm$0.04  &  -0.88$\pm$0.05  &  4.4$\pm$1.0\\ 
Sun  & 540 & 4550 & 1320 &  5771$\pm$1  &  4.44$\pm$0.00  &  0.02$\pm$0.05  &  1.6$\pm$0.3\\ 
$\tau$ Cet  & 570 & 1650 & 2150 &  5414$\pm$21  &  4.49$\pm$0.02  &  -0.50$\pm$0.03  &  0.4$\pm$0.4\\ 
\hline
\end{tabular}
\end{table*}

\section{Exploring the internal consistency of ESPRESSO spectra using solar data}           \label{sec:espresso}

High-precision stellar spectroscopy requires high-resolution and high S/N spectra. These HR spectra are typically taken with cross-dispersed echelle spectrographs, the reduction of which is difficult \citep{Piskunov-02, Prugniel-01}. In October and November 2018, we obtained five ESPRESSO UHR spectra of the Sun reflected from Vesta. The S/N of these spectra ranges from about 200 to 300. In this section we use these spectra at different stages of reduction and with different formats to study their effect on the spectral line properties.

\subsection{The ESPRESSO spectrograph}

ESPRESSO is a fibre-fed, cross-dispersed, high-resolution echelle spectrograph installed at the VLT (ESO).  The spectrograph was designed to achieve a 10 cm s$^{-1}$ -- level precision in radial velocity. 

\begin{figure}
\begin{center}
\begin{tabular}{c}
\includegraphics[angle=0,width=0.9\linewidth]{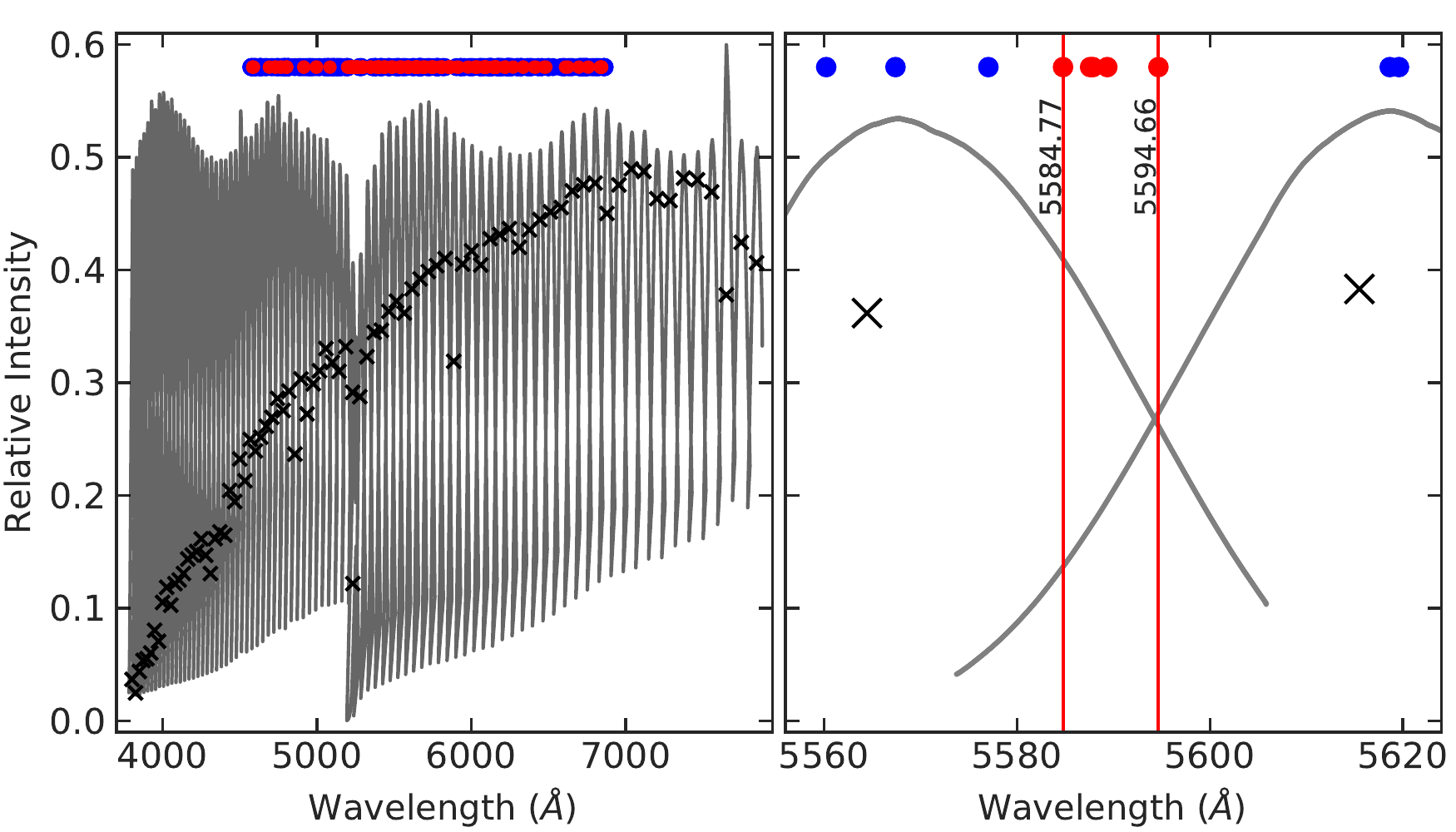}
\end{tabular}\end{center}
\vspace{-0.5cm}
\caption{Left panel: Spectral response (blaze function, black curve) for the ESPRESSO echelle orders for one of the solar spectra. The black crosses show the center of each order, and their ordinate values indicate the maximum S/N of the orders divided by 400 (for the sake of visibility). Blue and red circles indicate the wavelength of spectral lines that appear in single and multiple echelle orders, respectively. The two vertical red lines indicate the two spectral lines discussed in the text. Right panel: Zoom of the left panel for two adjacent spectral orders.}
\label{fig-orders_and_lines}
\end{figure}

The spectral resolution of ESPRESSO can be $\sim$70\,000 (multiMR mode), $\sim$140\,000 (singleHR mode), or up to $\sim$220\,000 (singleUHR) depending on the configuration. Because a pupil slicer (anamorphic pupil slicing unit) is used at the entrance of the spectrograph, ESPRESSO delivers two simultaneously observed spectra of the same target for the singleHR and singleUHR modes. The spectrograph produces 85 interference orders over two detectors, which together cover a wavelength range between 3782 and 7887\,\AA{}. Within a single interference order, the transmission variations are dominated by the blaze function\footnote{The blaze function is the chromatic variation of transmission and results form the interference patterns of the grooves of the echelle grating. For echelle gratings operating at high resolution, the small grove spacings leads to broad blaze functions that for efficiency, peak close to the center of the order and have minimum values at the edges.}. At the boundary wavelengths ($\sim$5250\,\AA{}) of the blue- and red-arm detectors the efficiency is slightly lower because of the dichroic that splits the light into the two arms (see Fig.~\ref{fig-orders_and_lines}). 

The final data products generated by the ESPRESSO Data Reduction Software (DRS) include an extracted blaze-corrected 2D spectrum (spectrum in the wavelength-order space,  called S2D in DRS) and a merged, blaze-corrected, rebinned 1D spectrum (called S1D in DRS). The S1D and S2D spectra are brought to the baricenter of the Solar System by applying the Earth radial velocity measured along the line of sight at the time of the observation to the wavelength solution. For a detailed description of the spectrograph and its performance, we refer to Pepe et al. (2020, submitted). Further information about the spectrograph is available in the ESPRESSO User Manual documents\footnote{\url{https://www.eso.org/sci/facilities/paranal/instruments/espresso/doc.html}}.

\begin{figure}
\begin{center}
\begin{tabular}{c}
\includegraphics[angle=0,width=0.9\linewidth]{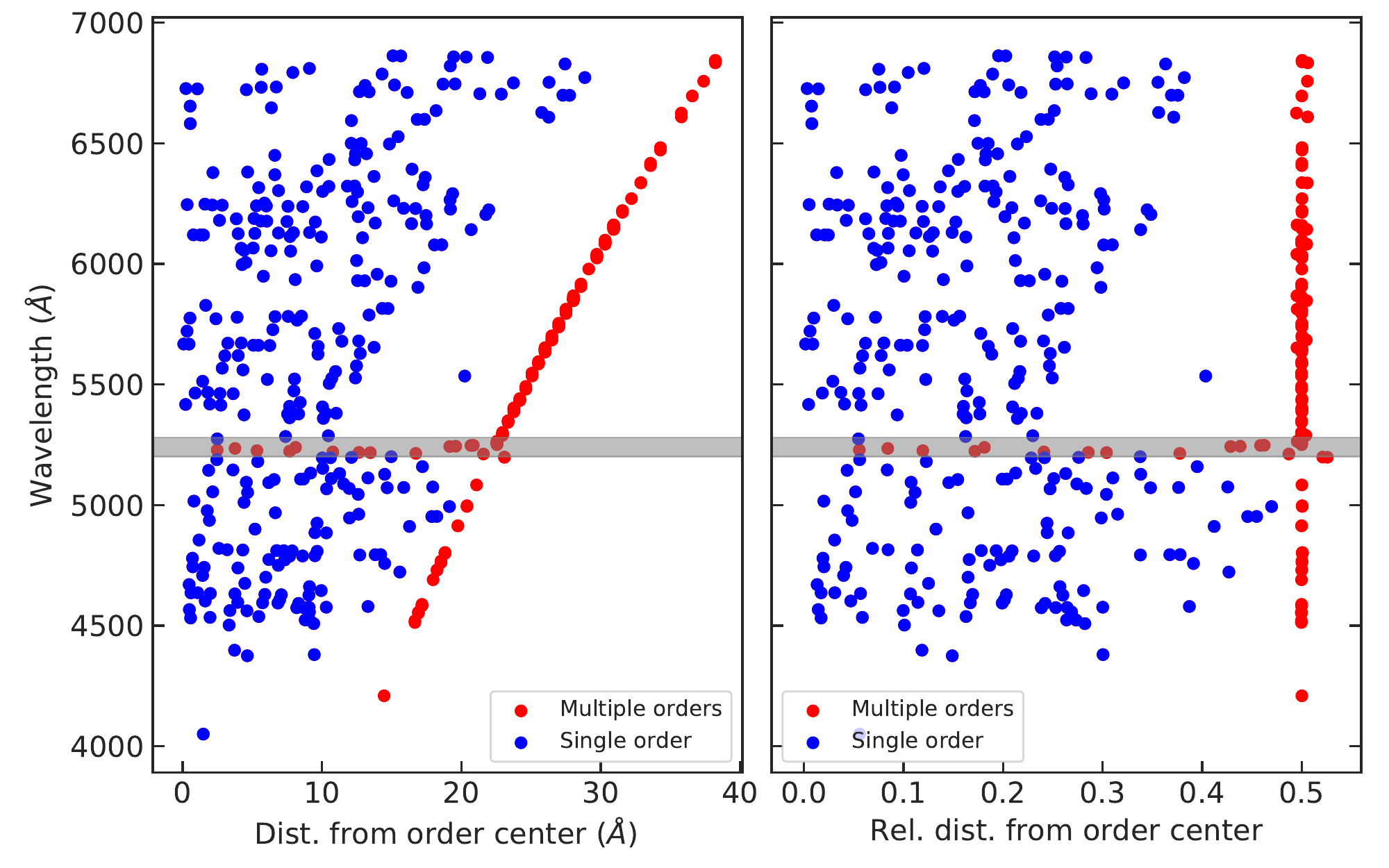}
\end{tabular}\end{center}
\vspace{-0.5cm}
\caption{Absolute (left panel) and relative (normalized to the size of the echelle order) distance (right panel) of the spectral lines from the center of the  ESPRESSO spectral orders. The
average distance is shown for lines observed in multiple orders. The boundary wavelength region (at  $\sim$5250\,\AA{}) of the blue- and red-arm detectors is indicated by the rectangle. }
\label{fig-lines_order_distance}
\end{figure}

\subsection{Lines in multiple spectral orders}                           \label{sec:multi_order}

ESPRESSO, like many multiple-order echelle spectrographs, has contiguous orders that cover the same wavelength range. Because of this overlap, some spectral lines appear in multiple spectral orders. It is very common to observe differences of up to a few percent in the line parameters observed in adjacent orders for echelle spectrographs \citep[e.g.,][]{Suzuki-03, Skoda-03, Hensberge-04}. Different approaches are proposed to solve the inconsistencies when the spectral orders are merged \citep[e.g.,][]{Erspamer-02, Hensberge-07, Brahm-17}. In this section we compare the spectral line parameters as measured in multiple orders of the 2D spectra.

In Fig.~\ref{fig-lines_order_distance}  we show the distance of the selected spectral lines from the centers of the spectral orders. For the lines observed in multiple orders, the average distance to the respective order's center is presented. We note that for lines observed in two orders, the relative average distance from the order centers is 0.5, that is,  half the distance (in angstroms) between the centers of two adjacent orders. About 35\% of the lines are observed in two orders, and 64\% are only observed in single orders. About 1\% of the lines with wavelengths close to the boundary wavelength ($\sim$5250\,\AA{}) of the blue and red arms appear in three orders. The overlap in wavelength coverage in orders is maximized to guarantee full coverage at the cutoff wavelengths. The figure shows that lines in single orders have shorter distances from the center of the orders than their counterparts observed in multiple orders. 
Because in an echelle spectrograph the wavelength coverage per order increases with wavelength, the average distance increases in the same way.

\begin{figure*}
\begin{center}
\begin{tabular}{cc}
\includegraphics[angle=0,width=0.45\linewidth]{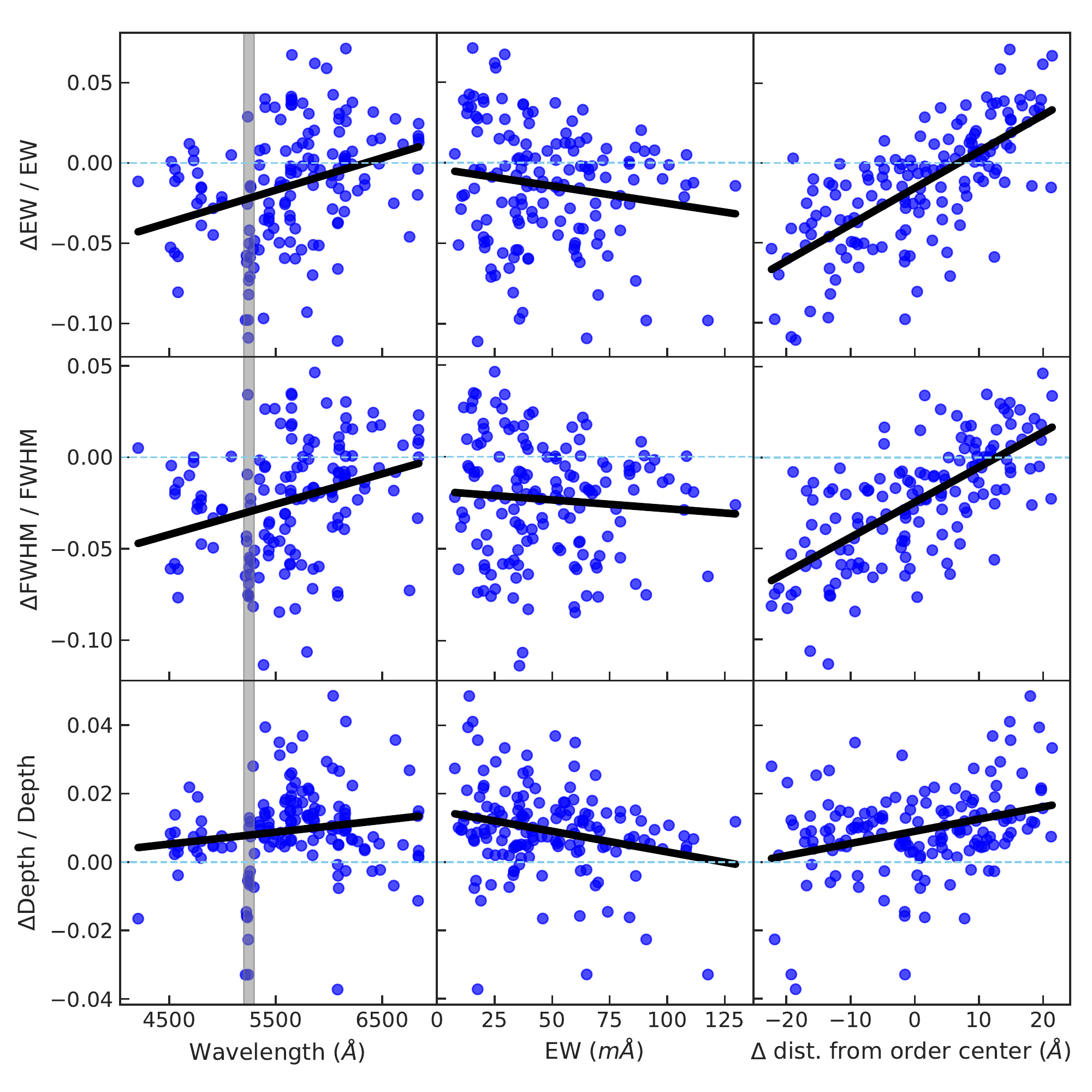} &
\includegraphics[angle=0,width=0.45\linewidth]{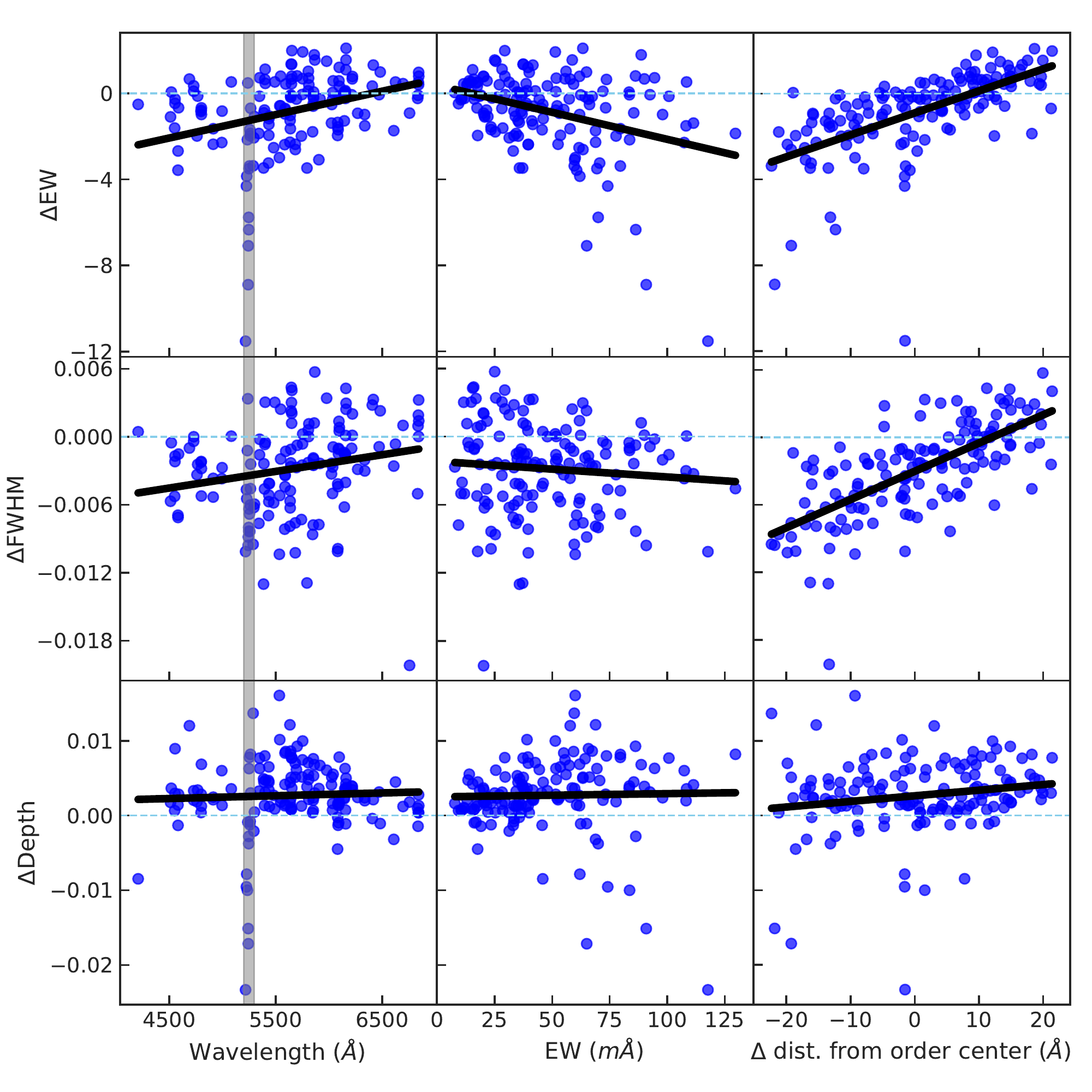}
\end{tabular}\end{center}
\vspace{-0.5cm}
\caption{Relative (left panel) and absolute (right panel) differences of spectral line parameters measured in different ESPRESSO echelle orders as a function of wavelength, EW, and the distance from the center of the spectral order. The boundary wavelength region (at  $\sim$5250\,\AA{}) of the blue- and red-arm detectors is indicated by a gray rectangle. Line properties have been measured with ARES with a fixed S/N for each order. The linear regression of the data is shown as black lines.}
\label{fig-sun_multiple_order}
\end{figure*}

In Fig.~\ref{fig-sun_multiple_order} we show how the relative and absolute differences of spectral line parameters (EW; line depth; and full width at half maximum, FWHM) as measured in different echelle orders depend on wavelength, EW, and the difference in distance from the centers of the spectral orders (``$\Delta$ dist. from orders centers''). This difference in distances from order centers, as the name suggests, is calculated as the difference of the spectral line distances from the adjacent order centers. For example, if its value is zero, then the line is located at the same distance from the centers of the two orders. A negative value means that the distance of the line from the longer wavelength order center is larger than its distance from the shorter wavelength order center (e.g., the 5584.77\,\AA{} line in Fig.~\ref{fig-orders_and_lines}). 

The relative differences of the line characteristics were calculated in the following way. For each solar spectrum and for each spectral order, we measured the line parameters with ARES. Then for each line observed in multiple orders we calculated the relative difference (difference divided by the mean) in the line parameter. As a final relative difference for each spectral line parameter, we used the average value of differences obtained for the five solar spectra. Lines with very large relative differences (more than 3$\sigma$ different from the mean values calculated for all the lines) in parameters were considered as outliers and removed from the analysis. One of the main sources for the inaccurate line measurements (outliers) are telluric lines.   We note that to measure the spectral line parameters, ARES requires the S/N for the spectrum (or in this case, the S/N for each spectral order) to estimate the noise and place the continuum. The S/N at the center of each order was extracted from the headers of the S2D spectra. 

The mean (and the standard deviation around the mean) relative difference for EW, FWHM, and line depth calculated for all the spectral lines is -1.4$\pm$3.6\%, -2.3$\pm$3.1\%, and 0.9$\pm$1.2\%, respectively. While these average differences are smaller than the scatter around these values, the left panel of Fig.~\ref{fig-sun_multiple_order} shows a clear dependence of $\Delta$EW/EW and $\Delta$FWHM/FWHM on the $\Delta$ distance from order centers, suggesting that the measurements of the line properties depend on the location of the lines relative to the center of the order. Moreover, the fact that linear regression fits do not cross the zero points when the $\Delta$ distance from order centers is zero suggests that the asymmetry of the blaze function plays a role.

Blazed gratings do not provide a constant efficiency over wavelength. As a consequence of this efficiency distribution, the S/N decreases at the order edges \citep{Eversberg2015}. Consequently, the spectral lines that appear in two adjacent orders will be observed at different S/N. This is best illustrated in the right panel of Fig.~\ref{fig-orders_and_lines}, where the 5594.66\,\AA{} line is observed at the same S/N in both orders, while the S/N at the 5584.77\,\AA{} line is about  $\sqrt{3}$ times (S/N goes with square root of flux) higher in the left order than in the right. Because in our line parameter measurements with ARES we considered a single value (maximum value which corresponds to the center of the order) of S/N for each spectral order the accuracy of the continuum placement might decrease with the distance from the order center.

\begin{figure*}
\begin{center}
\begin{tabular}{cc}
\includegraphics[angle=0,width=0.45\linewidth]{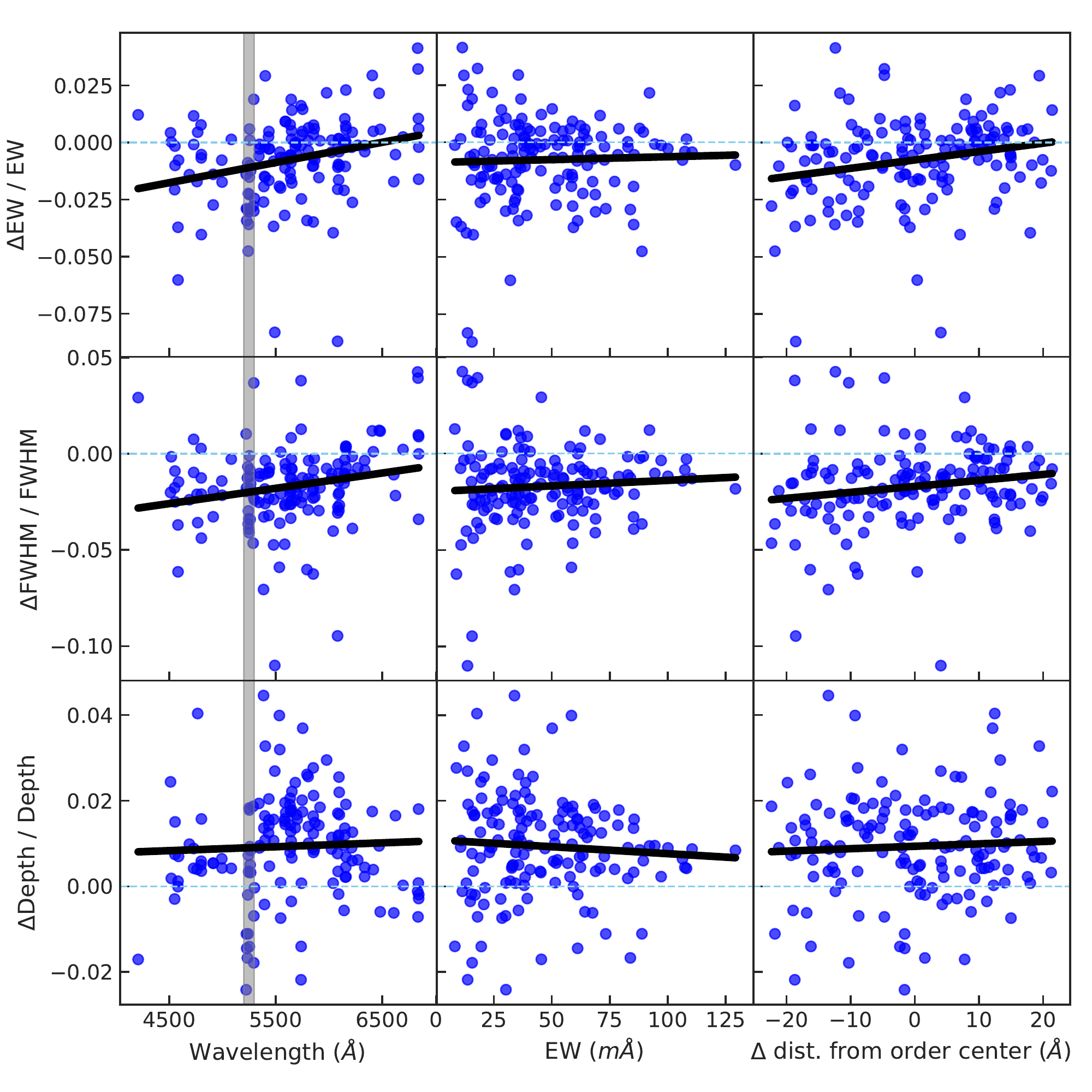} &
\includegraphics[angle=0,width=0.45\linewidth]{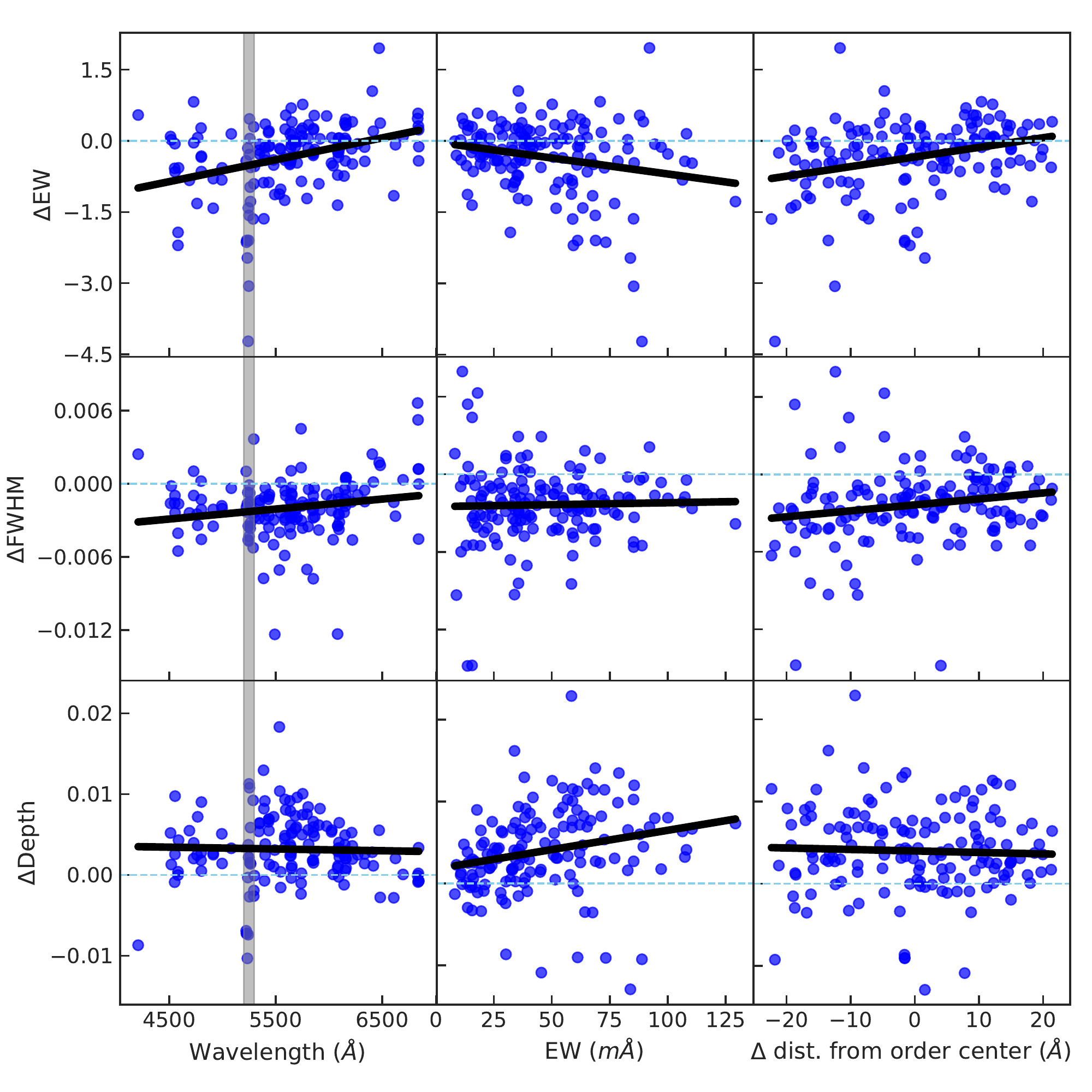}
\end{tabular}\end{center}
\vspace{-0.5cm}
\caption{Same as Fig.~\ref{fig-sun_multiple_order}, but for a S/N that varies as a function on the position of the spectral line in the echelle order. See the text for details.}
\label{fig-sun_multiple_order_varSNR}
\end{figure*}

We decided to repeat the line measurements with ARES, but varied the S/N for each line according to their distance from the center of the order. The approximate S/N around each line was calculated from the flux (photon counts) at the nearby continuum. The results of the test are shown in Fig.~\ref{fig-sun_multiple_order_varSNR}. It is apparent that the trend with distance from the order center almost disappeared. It is also important to note that the mean relative (and absolute) differences and their standard deviations decreased significantly and became -0.7$\pm$1.8\%, -1.7$\pm$2.0\%, and 0.9$\pm$1.1\% for EW, FWHM, and line depth, respectively. This decrease is especially apparent for the absolute value of the difference obtained for the EW, which changed from -0.8$\pm$1.9 to -0.3$\pm$0.7 m\AA{}.

\begin{figure}
\begin{center}
\begin{tabular}{c}
\includegraphics[angle=0,width=1\linewidth]{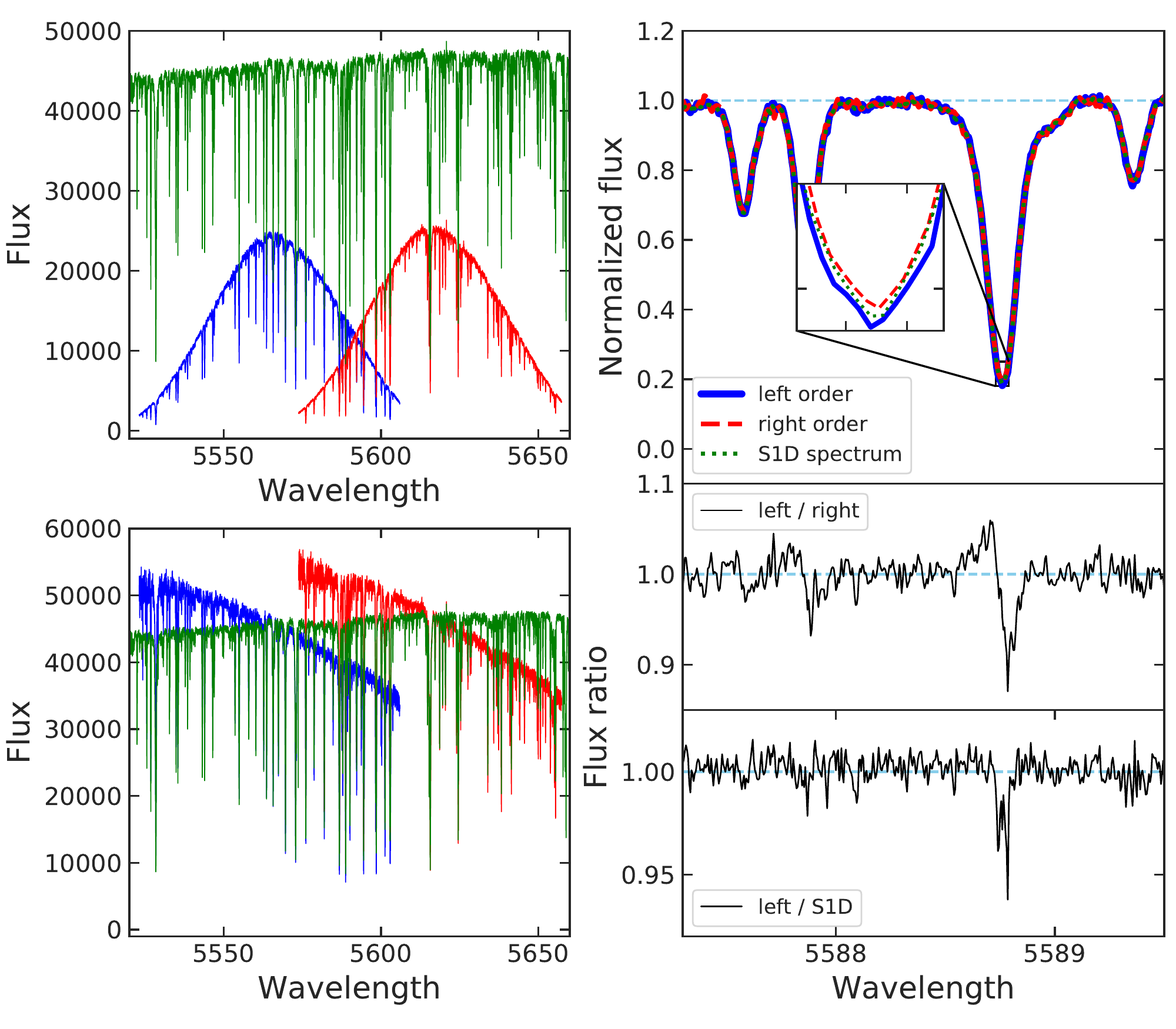}
\end{tabular}\end{center}
\vspace{-0.5cm}
\caption{Comparison of S1D spectrum with two ESPRESSO echelle orders before (top left panel) and after (bottom left panel) blaze correction. The top right panel compares the normalized S1D spectrum with the normalized S2D adjacent orders for a small spectral region in which the two orders overlap. The inset plot shows the zoom of the core of the strongest spectral line in the region. The middle-right and bottom-right panels show the division of the two orders and the division of the left order by the S1D spectrum, respectively. }
\label{fig-sun_orders_example_lines}
\end{figure}

Despite the improvement of the line parameter measurements with variable S/N, Fig.~\ref{fig-sun_multiple_order_varSNR} shows that an offset still remains (statistically insignificant, however) in FWHM and line depth measurements for the lines measured in two adjacent orders. In the top right panel of Fig.~\ref{fig-sun_orders_example_lines} we show a small spectral region around 5588\,\AA{} in which we compare the normalized spectra of  two spectral orders. The inset plot shows the core of the line, demonstrating that the line observed in the left (short wavelength) order is deeper than in the right (long wavelength) order. This is more clearly demonstrated in the middle right panel, which shows the result of the division of the two normalized spectra. The origin of this difference might be related to the widths of the lines being slightly different in the consecutive echelle orders. Because the spectral resolution within each echelle order increases with wavelength (Pepe et al. 2020, submitted), the same line located in the overlapping region of two adjacent orders will be observed at higher resolution in the short-wavelength order when compared with the longer-wavelength order \footnote{Because of varying resolution and sampling in spectral orders, we had to resample  the two spectral orders to a common fixed step size to produce the middle right panel of Fig.~\ref{fig-sun_orders_example_lines}.}.

The tests we performed in this section suggest that the spectral lines measured automatically with ARES from different spectral orders can provide consistent results, well within the uncertainties, if the S/N is correctly measured around each spectral line. However, for some spectral lines, a difference of up to $\sim$4-5\% in the EW can be observed. Figs.~\ref{fig-sun_multiple_order} and ~\ref{fig-sun_multiple_order_varSNR} suggest that the largest differences are observed for the lines that are located close to the boundary of the blue and red arms at $\sim$5250\,\AA{}.

\subsection{S1D vs. S2D spectra}                                        \label{sec:1d_vs_2d}

In this section we compare the spectral line parameters as measured from the S1D  and S2D spectra. A 1D spectrum is the final product of ESPRESSO DRS, which is created by merging the rebinned and blaze-corrected echelle orders and coadding the two slices. In the left panels of Fig.~\ref{fig-sun_orders_example_lines} we show a small portion of the S1D spectrum together with the two adjacent echelle orders from the 2D spectra before (left top) and after (left bottom) correction for the blaze function.

\begin{figure*}
\begin{center}
\begin{tabular}{cc}
\includegraphics[angle=0,width=0.45\linewidth]{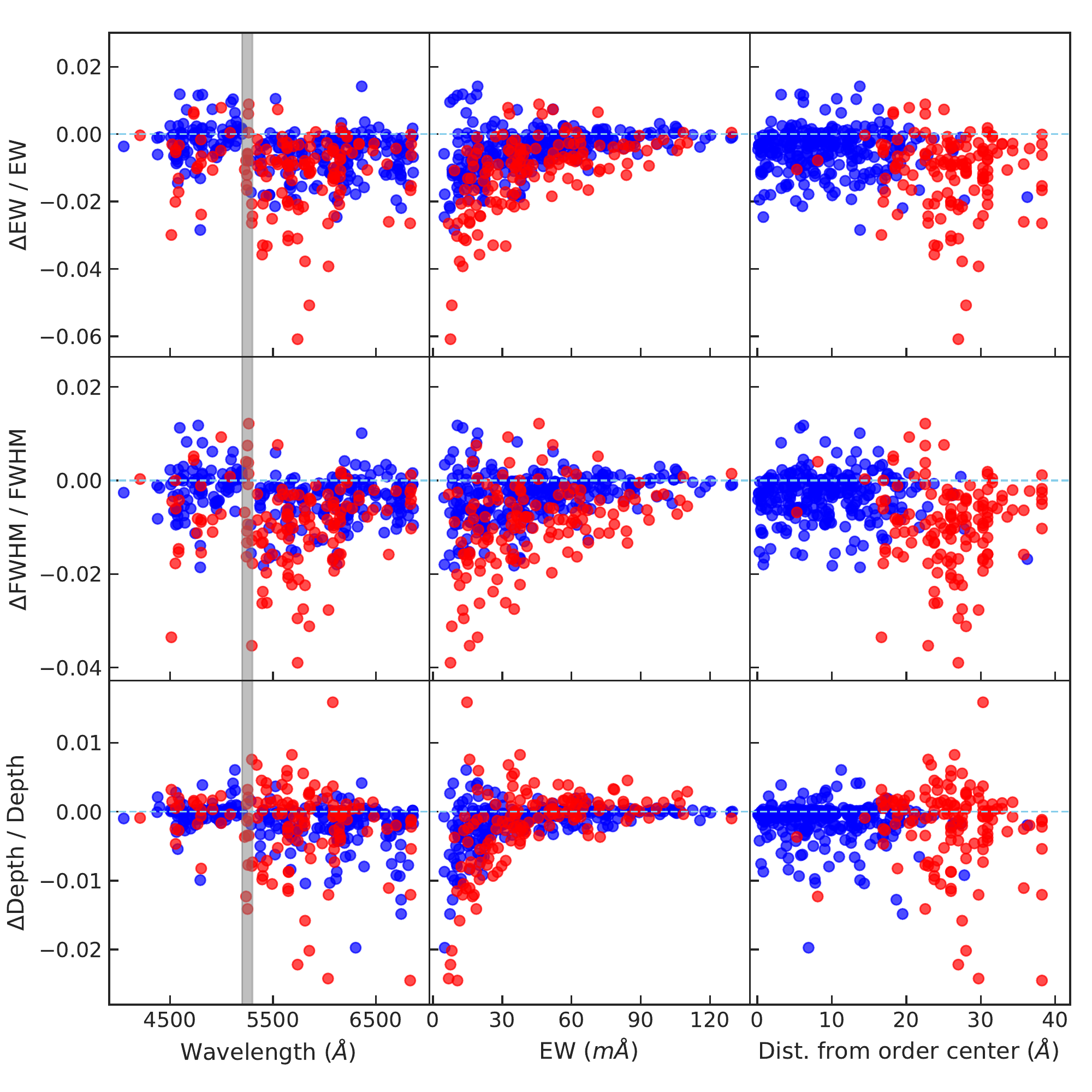} &
\includegraphics[angle=0,width=0.45\linewidth]{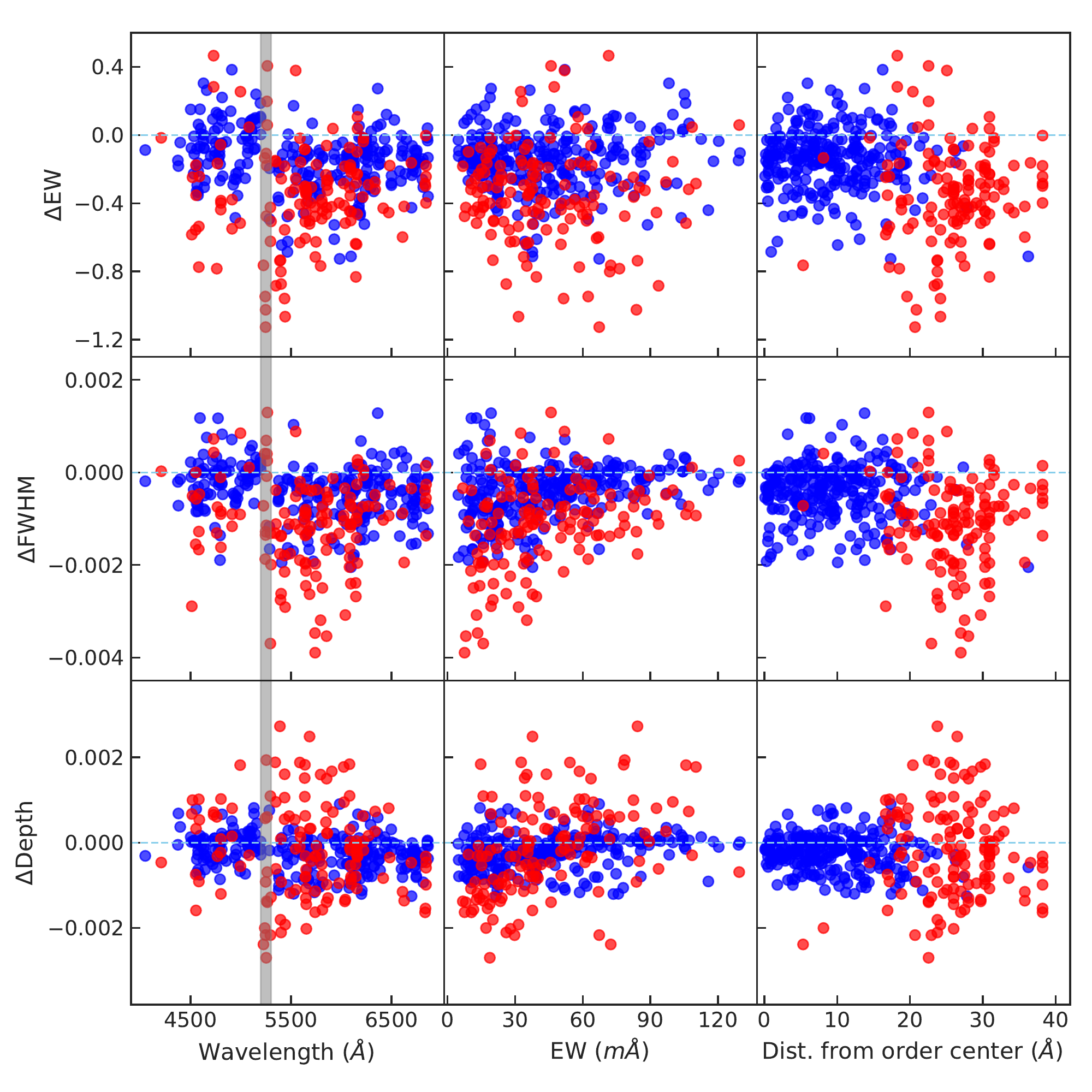}
\end{tabular}\end{center}
\vspace{-0.5cm}
\caption{Relative (left panel) and absolute (right panel) differences of the spectral line parameters measured from 1D and 2D spectra as a function of wavelength, EW, and the distance from the center of the spectral order. The red and blue symbols correspond to the lines observed in multiple orders and in single orders, respectively. The boundary wavelength region (at  $\sim$5250\,\AA{}) of the blue- and red-arm detectors is indicated by the rectangle. Line parameters for 2D spectra have been measured with ARES with a S/N that varies as a function of the position of the spectral line in the echelle order.}
\label{fig-sun_s1d_s2d_variable_snr}
\end{figure*}

The line measurements for 2D spectra were made as in the previous section by considering variable S/N depending on the position of the spectral lines. For each spectral line, the average values of the measurements from the two slices were taken. For the line measurements from S1D spectra, we considered a different S/N for each spectral order as extracted from the headers of the S1D fits files. The results of our measurements are shown in Fig.~\ref{fig-sun_s1d_s2d_variable_snr}. Again the average values obtained for the five solar spectra are shown. The outlier lines (at a 3$\sigma$ level) were removed. For the spectral lines observed in different orders, the average values are considered.

In general, we found relatively good agreement between the line parameter measurements. The average relative difference in EW, for example, is -0.6$\pm$0.8\% and the absolute difference is -0.19$\pm$0.22 m\AA{}. The relative differences in line parameters further decrease for relatively strong lines with EW $>$ 30  m\AA{}. However, Fig.~\ref{fig-sun_s1d_s2d_variable_snr} shows that the discrepancy between the line parameter measurements is larger for the lines that are positioned in multiple orders. In particular, the relative difference in EW  is larger by a factor of two for lines in multiple orders ($\Delta$EW/EW = -1.1$\pm$1.1\%) than for the lines observed in single orders ($\Delta$EW/EW = -0.5$\pm$0.6\%). A similar behavior is also observed for the FWHM of the lines where $\Delta$FWHM/FWHM increases from -0.3$\pm$0.5\% for  single lines to -0.9$\pm$0.8\% for  multiple lines.

In Sect.~\ref{sec:multi_order} we proposed that becaue of the slight variation in spectral resolution from the centers to the edges of spectral orders, the widths of spectral lines appear different in different spectral orders. In the top right panel of Fig.~\ref{fig-sun_orders_example_lines} we compare the normalized spectra of two adjacent orders with the S1D spectrum for a small spectral region. The inset plot shows that the depth of the line in the S1D spectrum is intermediate between the lines  observed in the two adjacent orders. The bottom right panel shows the residuals of the division of the left (short wavelength) 2D order by the 1D spectrum.   
These results suggest that the observed difference in FWHM and EW for the lines measured from 1D and 2D spectra is probably related to the slight variation in spectral resolution from the centers to the edges of spectral orders.

\begin{figure}
\begin{center}
\begin{tabular}{c}
\includegraphics[angle=0,width=0.9\linewidth]{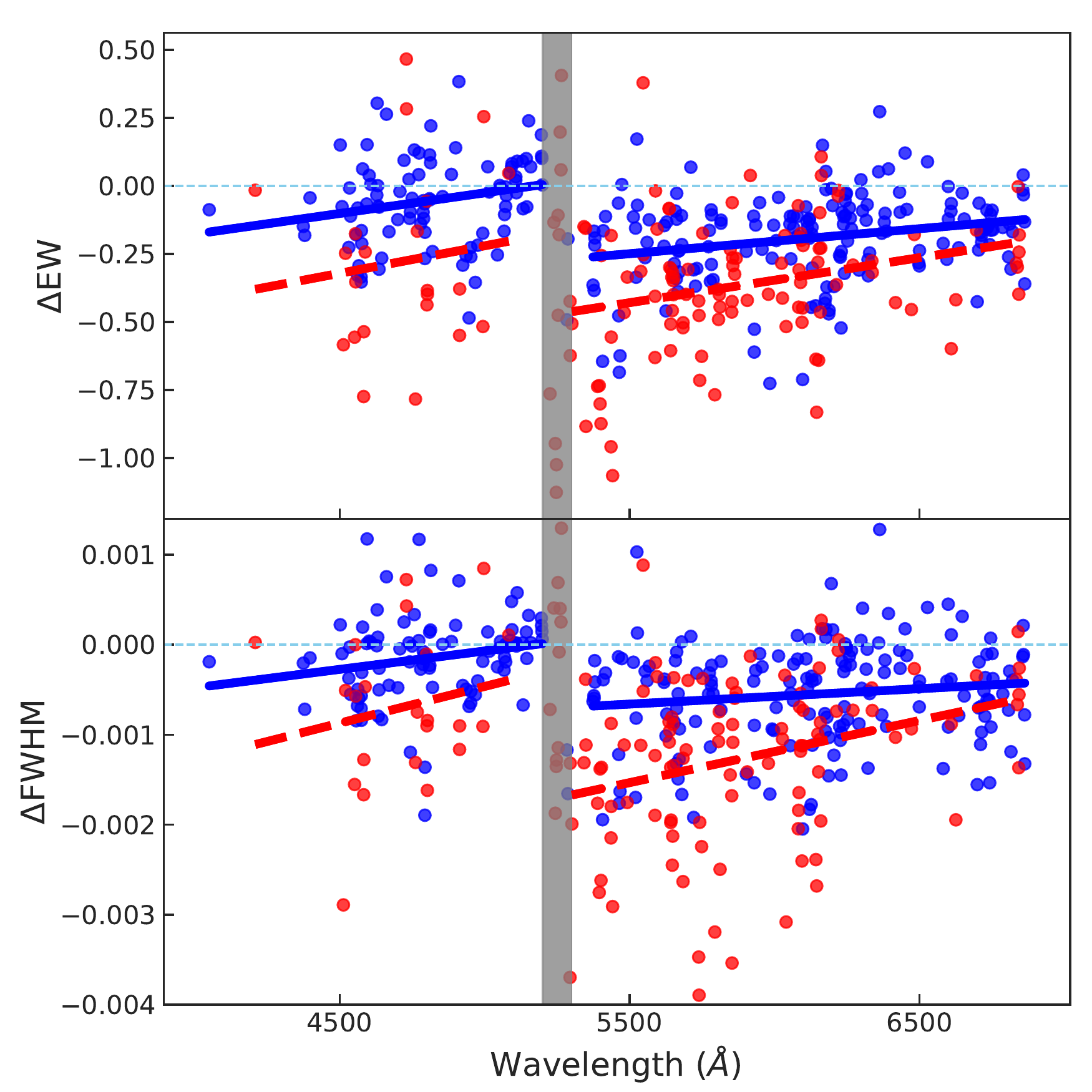}
\end{tabular}\end{center}
\vspace{-0.5cm}
\caption{Same as the right panel of Fig.~\ref{fig-sun_s1d_s2d_variable_snr}, but only for $\Delta$EW and $\Delta$FWHM. The two blue lines show the results of linear regression of the lines in single orders located in the blue and red arms. The linear fits for the lines in multiple orders in the blue and red arms are shown as red dashed lines.}
\label{fig-sun_rel_s1d_s2d_variable_snr_EW}
\end{figure}

\begin{figure}
\begin{center}
\begin{tabular}{c}
\includegraphics[angle=0,width=0.9\linewidth]{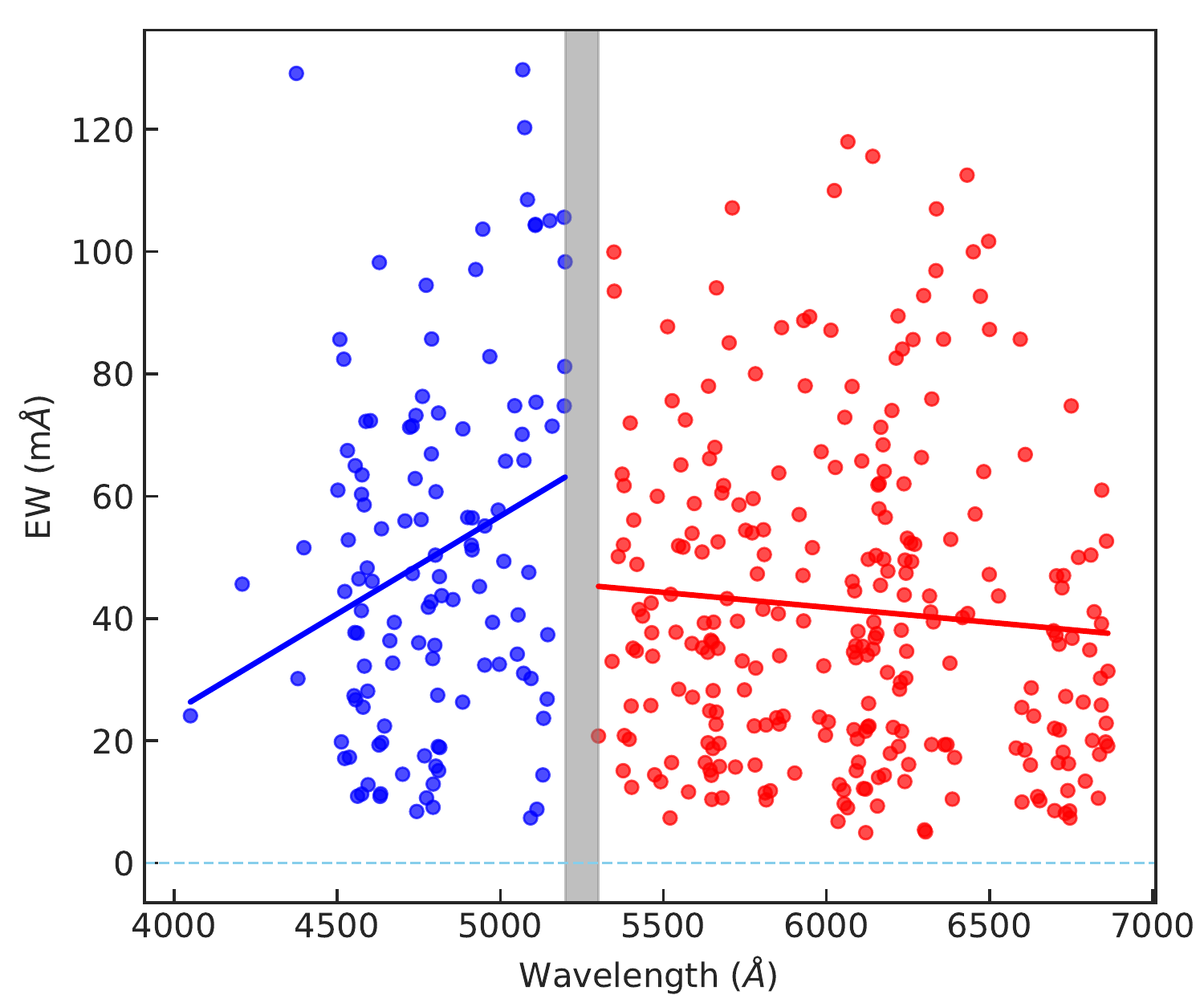}
\end{tabular}\end{center}
\vspace{-0.5cm}
\caption{Equivalent width of spectral lines measured from S1D spectra as a function of wavelength. Spectral lines in the blue- and red-arms of the spectrograph are represented by blue and red symbols, respectively. The results of linear regression for the two sets of lines are shown as solid blue and red lines.}
\label{fig-ew_wavelength}
\end{figure}

Another interesting feature is visible in Fig.~\ref{fig-sun_s1d_s2d_variable_snr}. There seems to be an offset in EW and FWHM differences between lines located in the blue and red arms. In Fig.~\ref{fig-sun_rel_s1d_s2d_variable_snr_EW} we show the dependence of $\Delta$EW and $\Delta$FWHM on wavelength, where the linear regression was performed for the lines located in the blue and red arms separately. The plot shows that the offset is visible both for lines located in only single orders and lines located in multiple orders. It is difficult to identify the exact reason of this offset. Because of the optical-design and manufacturing parameters, the spectral resolution at the red end of the blue arm is different from the resolution at the blue end of the red arm (Pepe et al. 2020, submitted). Our interpretation is that the observed trend is a result of the combination of the aforementioned difference in spectral resolution between the two arms, slight dependence of the $\Delta$EW and $\Delta$FWHM on EW (right panel of Fig.~\ref{fig-sun_s1d_s2d_variable_snr}), and the complex dependence of EW on wavelength (see Fig.~\ref{fig-ew_wavelength}). Fig.~\ref{fig-ew_wavelength} shows that the average EW of the spectral lines at the reddest end of the blue arm is larger than that of the shortest wavelength spectral lines of the red arm.

\begin{figure}
\begin{center}
\begin{tabular}{c}
\includegraphics[angle=0,width=0.9\linewidth]{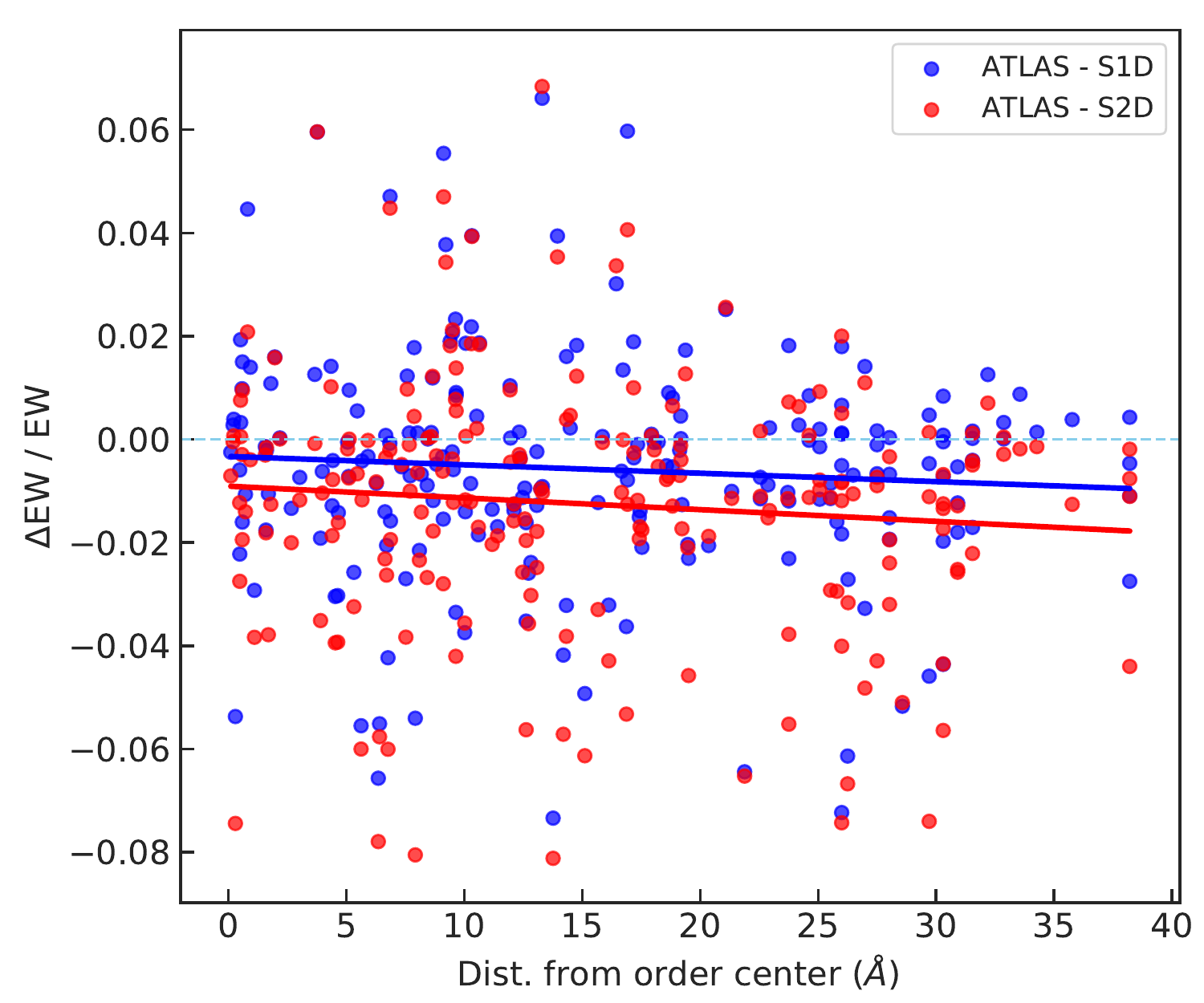}
\end{tabular}\end{center}
\vspace{-0.5cm}
\caption{Comparison of the EWs of spectral lines measured from S1D (blue symbols) and S2D (red symbols) with those of \citet{Sousa-08}. Outliers (at 5$\sigma$ level) are excluded from the plot.}
\label{fig-ew_comparision}
\end{figure}

Although the observed difference in EWs as measured from 2D and 1D spectra is smaller than $\sim$1\%, it is worth trying to understand which measurements are more accurate and therefore preferable. \citet{Sousa-08} measured EWs of about 300 iron lines  manually (using the IRAF\footnote{IRAF is distributed by National Optical Astronomy Observatories, operated by the Association of Universities for Research in Astronomy, Inc., under contract with the National Science Foundation, U.S.A.} ``splot'' routine)  from the Kurucz solar atlas \citep{Kurucz-05}\footnote{\url{http://kurucz.harvard.edu/sun.html}}. Assuming these measurements are accurate, we compared our EW measurements from 1D and 2D spectra with these  reference values. Our comparison (Fig.~\ref{fig-ew_comparision}) suggests better agreement for the measurements performed on the 1D spectra (-0.6$\pm$2.2\%) than on the 2D spectra (-1.2$\pm$2.4\%). For both measurements based on the 1D and 2D spectra the disagreement with the solar atlas observations seems to increase for the lines that are located farthest from the center of the echelle orders, although the trend is statistically not significant (F statistics suggest a p-value of 0.2 that the slope is significantly different from zero, a probability that is too high to provide confidence on the fit).

Interestingly, 1D spectra  single lines (lines that are located in a single echelle order in the 2D spectra) show a slightly worse agreement (-0.6$\pm$2.3\%) with ATLAS measurements than the lines that are located in multiple echelle orders in the 2D spectra (-0.5$\pm$1.8\%). The opposite trend is observed for the 2D spectra where the agreement for the  single lines is slightly better: -1.2$\pm$2.5\% vs -1.4$\pm$2.2\%. These differences are not statistically significant and need to be taken only as indications.

The observational results that the measurements based on 1D spectra are more accurate than those based on 2D spectra may suggest that a single measurement of spectral lines from combined spectra is preferable to averaging multiple measurements of spectral lines from different orders and slices. It is possible that when more sophisticated averaging techniques are used (e.g., giving different weights to line measurements based on the S/N around the line or based on the distance from the center of the echelle orders), the results will be improved for 2D spectra. However, this might be only relevant for very specific science cases where extremely high precision in line parameter measurements is required.

\subsection{Combined versus individual spectra}                    \label{sec:sun_params}

\begin{figure}
\begin{center}
\begin{tabular}{c}
\includegraphics[angle=0,width=0.9\linewidth]{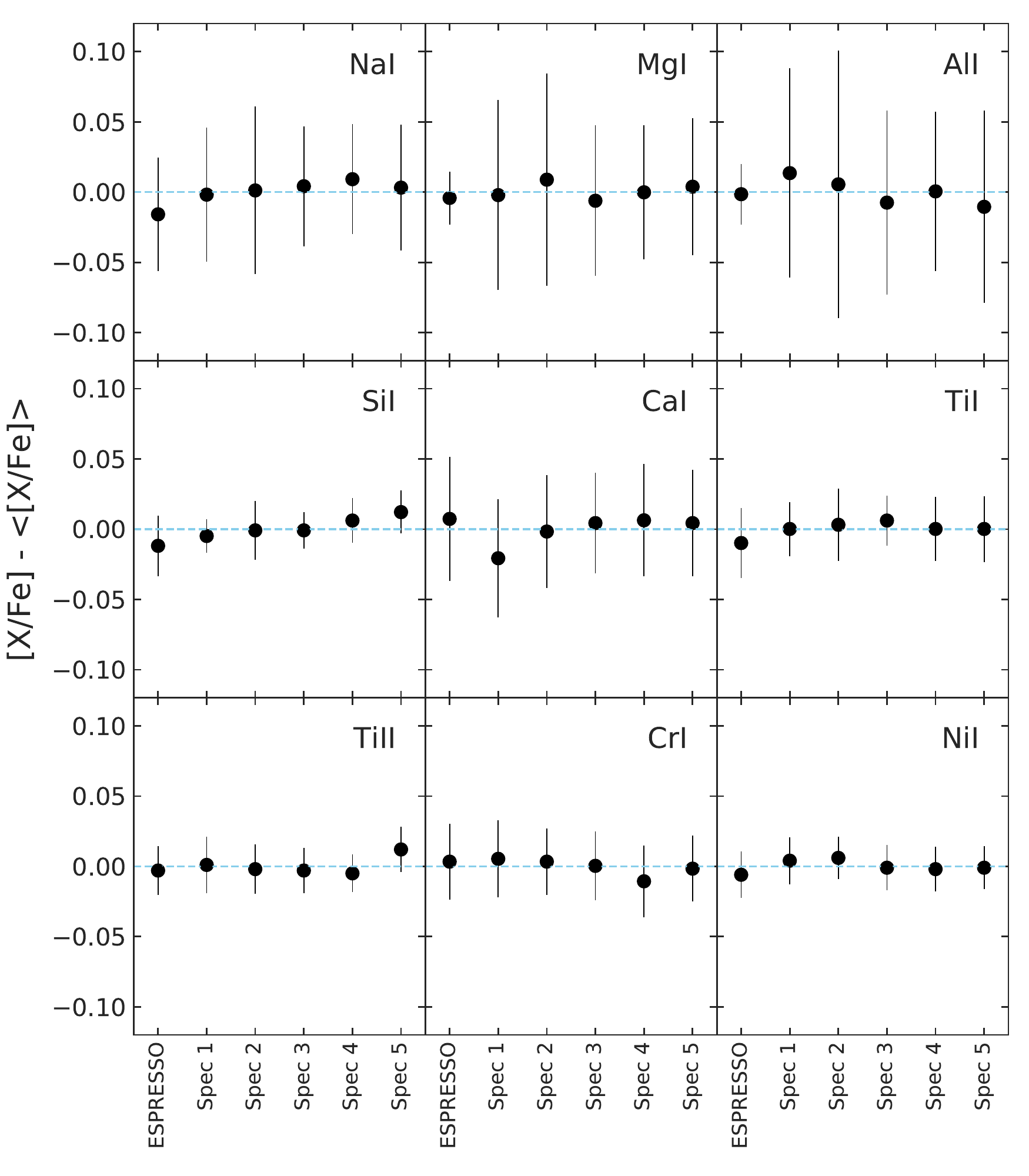}
\end{tabular}\end{center}
\vspace{-0.5cm}
\caption{[X/Fe] abundance ratios for some elements derived with the use of combined and S1D individual solar ESPRESSO spectra relative to the mean $<$[X/Fe]$>$.}
\label{fig-elfe_sun}
\end{figure}

In this section we compare the atmospheric parameters and abundances of several refractory elements derived from the combined ESPRESSO spectra with those derived from the five individual ESPRESSO S1D spectra. The choice of the elements was motivated by the choice of the EW method to determine their composition, that is, elements whose abundances require spectral synthesis or elements with lines that are significantly affected by the hyperfine splitting were not considered. The element abundances were derived as in \citet{Adibekyan-15a} using the same tools and atmosphere model as for the determination of the stellar parameters. Our results are presented in Table~\ref{tab:sun_parameters_abundances} and are shown in Fig.~\ref{fig-elfe_sun}. The figure shows [X/Fe] abundance ratios relative to the mean $<$[X/Fe]$>$ . The [X/Fe] instead of [X/H] is preferred to  compensate for the slightly different metallicities we obtained for each spectrum.

The table shows that the stellar parameters derived from the combined and individual spectra agree in very well. It might come as a surprise that the estimated uncertainties for the stellar parameters did not decrease for the combined spectra. However, we should note that the precision and accuracy of the stellar parameters depend not only on the quality of the spectra, but also on several other factors such as atomic data of the spectral lines and atmosphere models. When a differential spectroscopic analysis is performed, that is, when the atmospheric parameters of a star relative to another star are determined, the effect of the latter factors is reduced, which gives the greatest weight to the quality of the spectra \citep[][]{Saffe-16, Liu-20}. However, in our analysis we are interested in comparing the parameters and abundances determined from each spectrum individually. Additionally, when UHR spectra are combined, the process of combining (because of rebinning, e.g.) individual spectra also plays a role. 
Nevertheless, our results show that the determination of solar atmospheric parameters from the different spectra is robust at the level of $\sim$10 K for \teff, $\sim$0.03 dex for \logg, and $\sim$0.01 dex for [Fe/H].

Fig.~\ref{fig-elfe_sun} shows that the abundances of the elements also agree very well for the different spectra. The figure shows that the error bars for the elements (\ion{Na}{i}, \ion{Mg}{i}, and \ion{Al}{i}) with fewer than four spectral lines are significantly larger for the individual ESPRESSO spectra than the results obtained with the combined spectra. The main reason for this is that the methods of estimating the EW measurement uncertainties are different for lines with fewer or more than four spectral lines \citep[see][for details]{Adibekyan-16}. For the elements with more than three spectral lines (\ion{Si}{i}, \ion{Ca}{i}, \ion{Ti}{i}, \ion{Ti}{ii}, \ion{Cr}{i}, and \ion{Ni}{i} in our study), the line-to-line abundance scatter is used as an EW measurement error. For elements with three or fewer spectral lines (\ion{Na}{i}, \ion{Mg}{i}, and \ion{Al}{i} in our study) the errors on EWs were calculated following \citet{Cayrel-88}. These uncertainties take the statistical photometric error due to the noise in each pixel (as we discussed in Sect.~\ref{sec:1d_1d}) and the error related to the continuum placement into account, which is the dominant contribution to the error \citep[][]{Cayrel-88, BertrandeLis-15}. The larger error for the individual ESPRESSO spectra arises from the dependence of the uncertainty estimation by the formula on the S/N. Given the very good agreement in the abundance determination for \ion{Na}{i}, \ion{Mg}{i}, and \ion{Al}{i} (top three panels of Fig.~\ref{fig-elfe_sun}) based on different spectra, the EW measurement uncertainties might be overestimated.

\section{ESPRESSO versus  PEPSI versus HARPS}                                 \label{sec:spectrographs}

In this section we compare the line parameters and resulting stellar parameters of the 11 stars for which we have  ESPRESSO, PEPSI, and HARPS spectra. We measured the lines on the combined 1D spectra (more than one spectrum was available for most of the stars). Reduced S1D HARPS spectra were taken from the ESO Phase 3 archive\footnote{\url{http://archive.eso.org/wdb/wdb/adp/phase3\_main/form}}. The science-ready normalized 1D PEPSI spectra were taken from the PEPSI archive of ``Gaia benchmark stars and other M-K standards''\footnote{\url{https://pepsi.aip.de/?page\_id=552}}. As mentioned on the PEPSI webpage, the spectra were reduced with SDS4PEPSI Version 1.0, and its future releases may change the resulting spectra, in particular, the continuum setting of the very blue part. This should not be a problem for our analysis because ARES performs a local continuum normalization for each spectral line. Moreover, the spectral lines we used start at a wavelength of $\sim$4500 \AA, that is,{}  farther away by about 700 \AA{} from the blue wavelength limit of the PEPSI spectra. Because the available PEPSI spectra are already normalized, it is not possible to estimate the S/N from the flux. We therefore let ARES automatically determine and fix the S/N (at around 6000 \AA{}) for the spectra and place the continuum accordingly \citep[see][for more details about ARES]{Sousa-15}.

\begin{figure}
\begin{center}
\begin{tabular}{c}
\includegraphics[angle=0,width=0.9\linewidth]{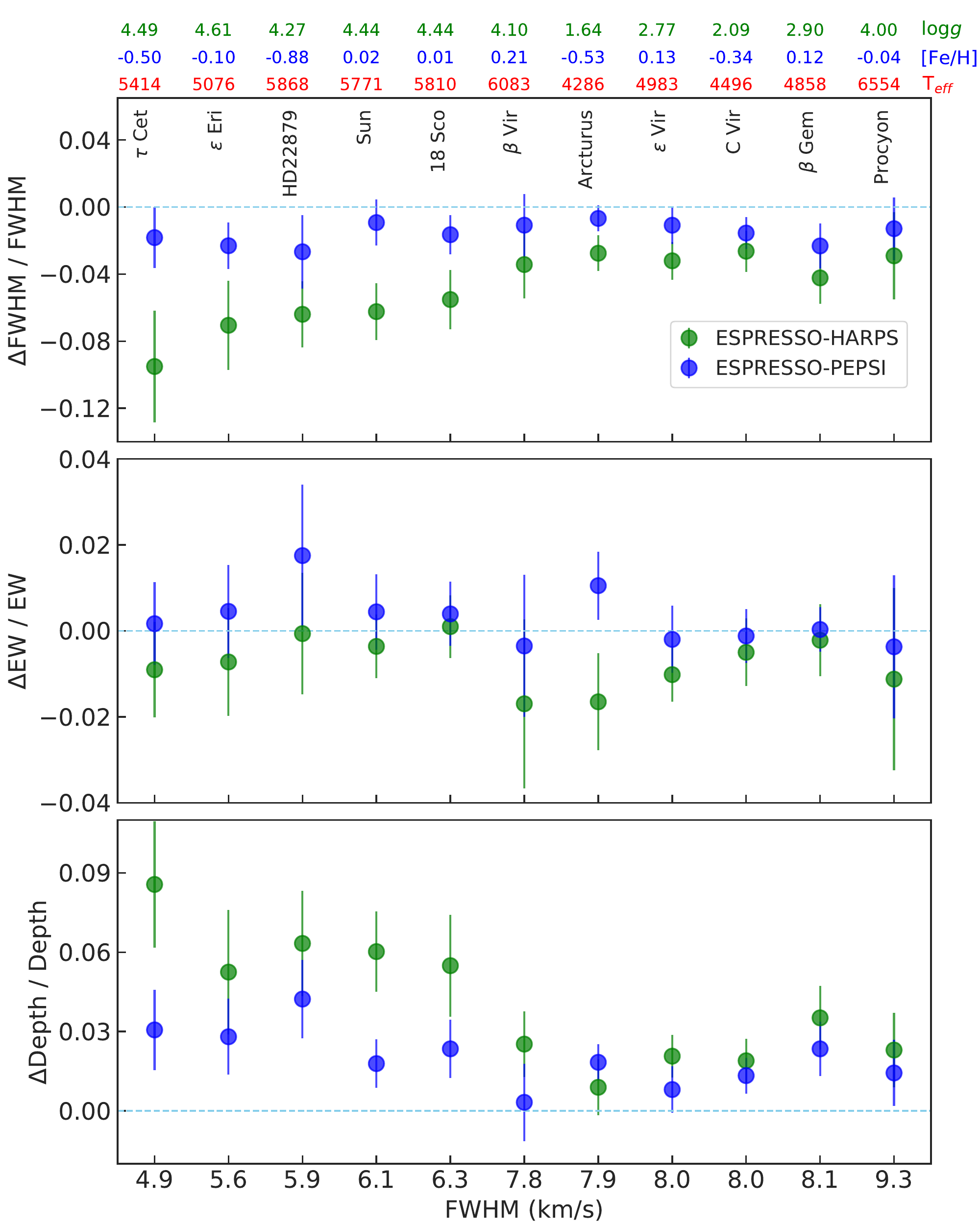}
\end{tabular}\end{center}
\vspace{-0.5cm}
\caption{Relative difference in line parameters between ESPRESSO, PEPSI, and HARPS as a function of the mean FWHM of the lines. The symbols correspond to the mean value calculated for all the spectral lines, and the error bar represents the line-to-line dispersion. The atmospheric parameters of the stars are shown at the top of the panels.}
\label{fig-EW_fwhm_depth}
\end{figure}

Fig.~\ref{fig-EW_fwhm_depth} shows the relative differences between the line parameter measurements from the ESPRESSO spectra and those of PEPSI and HARPS. As in the previous sections, the outlier measurements at the level of 3$\sigma$ were removed from the analysis. We plotted these differences as a function of the average FWHM of the spectral lines (measured from ESPRESSO spectra) because the effect of spectral resolution on line measurements should depend on the intrinsic broadening of the lines. In addition to the average width of the lines, the relative difference will also depend on other characteristics of the spectral lines such as on the average strength of the spectral lines. 

The bottom panel of the figure clearly shows that the lines appear deeper in ESPRESSO spectra than in PEPSI and HARPS. As expected, these differences are more pronounced for stars with intrinsically narrow spectral lines (small FWHM). For $\tau$ Cet, the slowest rotating star (Table~\ref{tab:all_parameters}), the average relative difference in line depth when ESPRESSO is compared with PEPSI and HARPS is $\sim$3$\pm$1.5\% and $\sim$8$\pm$2\%, respectively. For the fastest rotating star, this difference decreases to about 1\% when ESPRESSO is compared with PEPSI, and to about 2\% when ESPRESSO is compared with HARPS. 

\begin{figure*}
\begin{center}
\begin{tabular}{cc}  
\includegraphics[angle=0,width=0.4\linewidth]{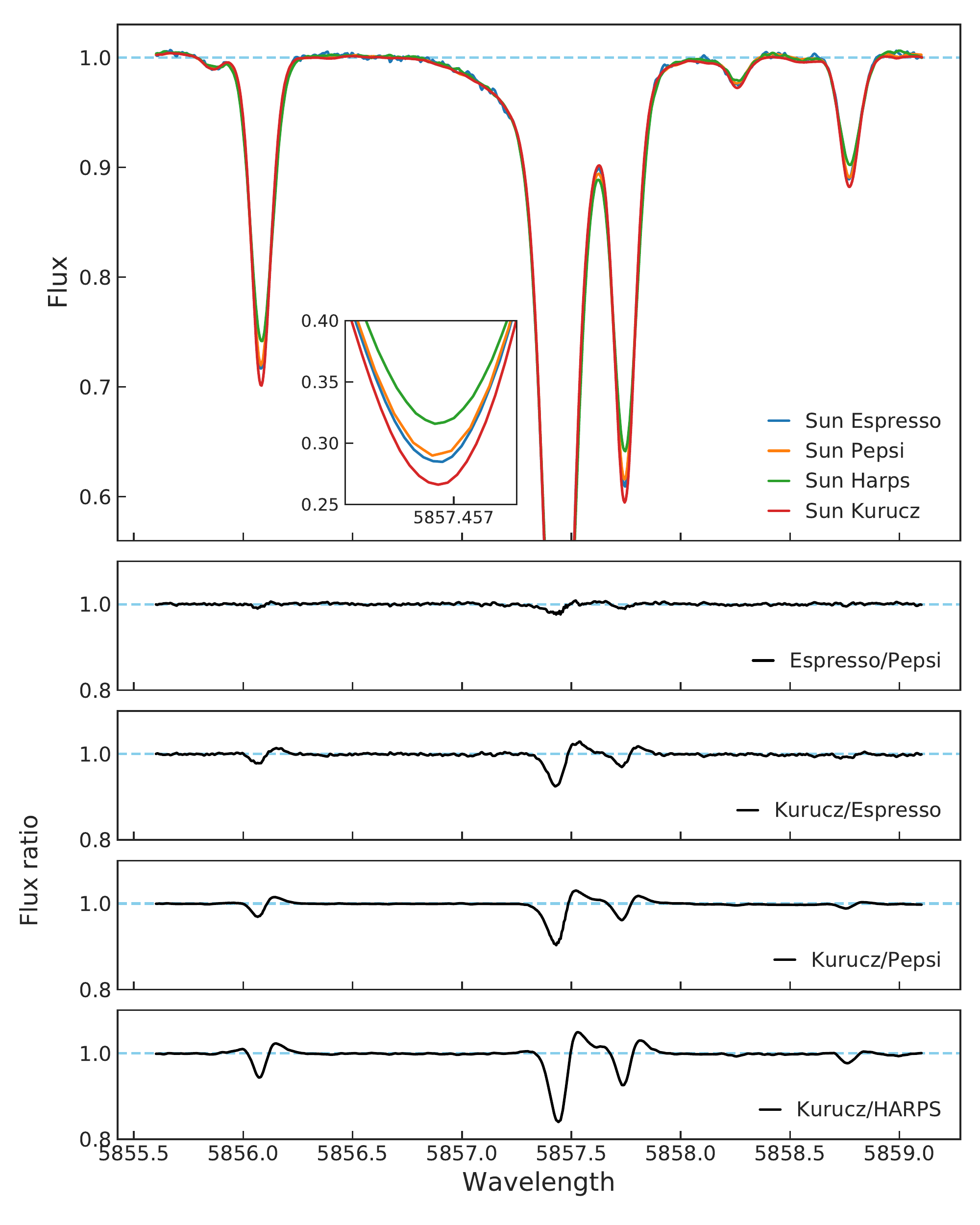} & 
\includegraphics[angle=0,width=0.4\linewidth]{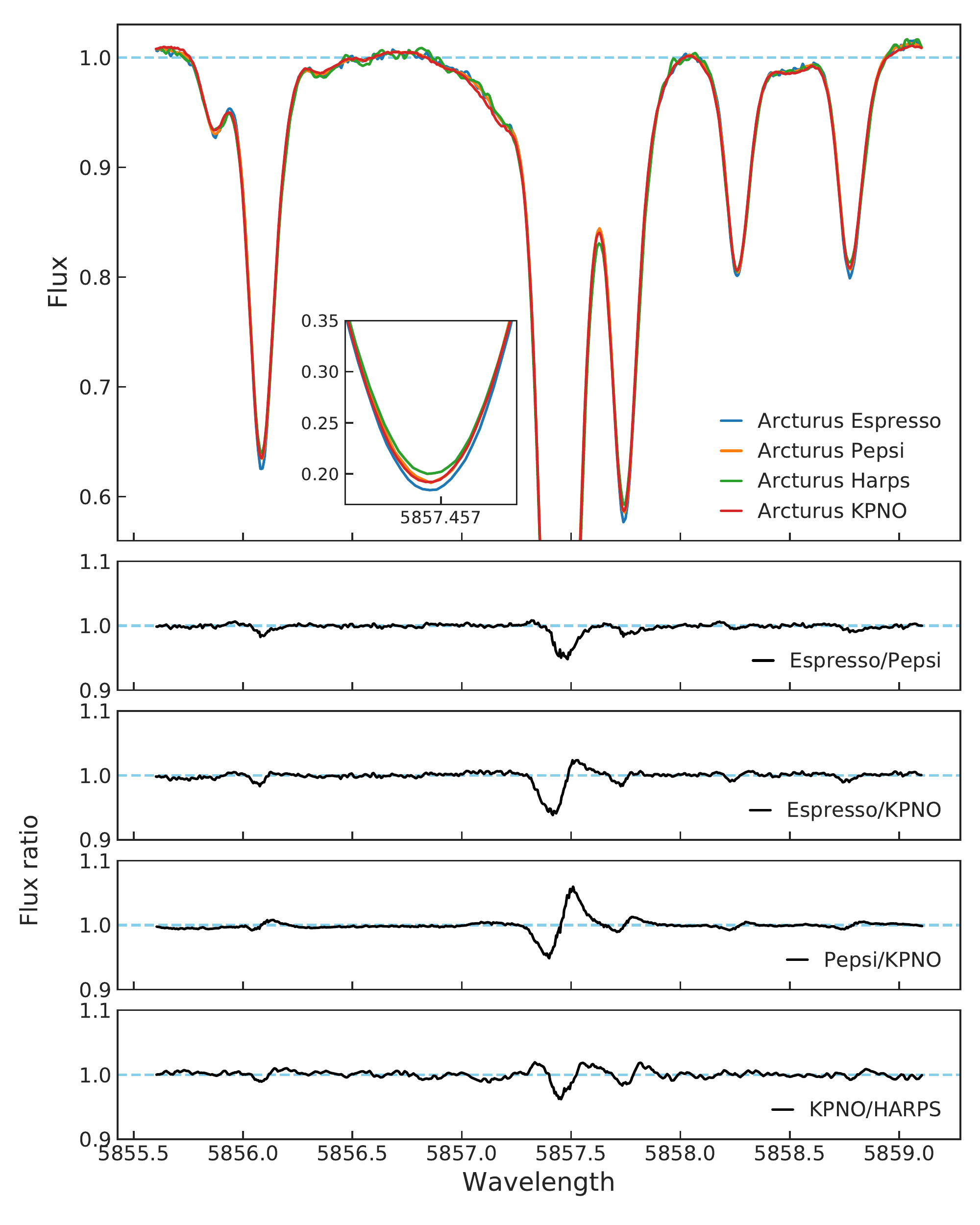}
\end{tabular}\end{center}
\vspace{-0.5cm}
\caption{Comparison of the ESPRESSO, PEPSI, and HARPS spectra of the Sun and Arcturus with the Kurucz solar atlas from \citet{Kurucz-05} and the KPNO Arcturus atlas from \citet{Hinkle-00}. The inset graphs show the core of the close-by strong lines. The four bottom panels show the ratio of the aforementioned spectra.}
\label{fig-sun_arcturus}
\end{figure*}

Because of the different spectral resolution of ESPRESSO, PEPSI, and HARPS ($\sim$220\,000, $\sim$220\,000, and $\sim$115\,000, respectively), and because the v$\sin i$ and other stellar broadening mechanisms are comparable to the resolution element of the spectrograph measured, the FWHM as measured on each spectrographs is expected to scale with resolution. However, it should be noted that the spectral samplings of the two spectrographs are different: 2.5 pixels and 2 pixels per resolution element for ESPRESSO and PEPSI, respectively. Additionally, the spectral resolution of the two spectrographs changes within the end of the echelle orders in different nonlinear ways \citet[][; Pepe et al. 2020, submitted]{Strassmeier-18}. Finally, on further investigation, it was recently found that the PEPSI spectra of the brightest stars (e.g., the Sun) used for comparison (taken before 2019) and available online were obtained with exposures times of about a few seconds. For these exposure times, CCD controler and shutter issues led to the flashing of white light on the detector, leading to shallower lines. These issues were solved for recently acquired spectra, and a new dataset, corrected for this effect, will soon be released to the community (Strassmeier 2020, priv. comm).

Fig.~\ref{fig-sun_arcturus} shows a comparison of the ESPRESSO, PEPSI, and HARPS spectra of the Sun and Arcturus for a small spectral window. For comparison, the Kurucz solar atlas (spectral resolution of about 500\,000) from \citet{Kurucz-05} and the Kitt Peak National Observatory (KPNO) Arcturus atlas (spectral resolution of about 150\,000) from \citet{Hinkle-00} are also shown. The plot shows an almost perfect match for the continuum between the different spectra. It also clearly shows (especially for the strongest lines) that the lines are indeed deeper in ESPRESSO spectra than in PEPSI and HARPS. As expected from its highest spectral resoling power, the deepest lines  for the Sun  are observed in the Kurucz atlas.  

The top panels of Fig.~\ref{fig-EW_fwhm_depth} show the difference in EWs and FWHM of the lines as measured with ARES from the ESPRESSO, PEPSI, and HARPS spectra. Because ARES assumes a Gaussian profile of the spectral lines when the fit is made, and because of the differences in line depths outlined in the previous sections, we expect to see (on average) higher values of the FWHM for the lines observed in HARPS and PEPSI spectra when compared to the ESPRESSO ones. The spectral lines that appear deeper in ESPRESSO spectra are also narrower. The maximum difference, observed for the slowest rotating star, is $\sim$9$\pm$2\%  and  $\sim$2$\pm$1\% between ESPRESSO and HARPS, and between ESPRESSO and PEPSI, respectively. 

While we see differences in the spectral line profiles (width and depth) measured from the ESPRESSO, PEPSI, and HARPS spectra, the middle panel of Fig.~\ref{fig-EW_fwhm_depth} shows that the EWs of the lines are practically conserved. In general, it can be assumed that for a conceptual spectrograph, the EWs of spectral lines are conserved and in particular are not a function of resolution (as long as the line is unblended). The average EW of the lines measured from the spectra of these spectrographs are indistinguishable within the error bars. The only exception is Arcturus, for which the difference is slightly above 1$\sigma$ ($\sim$-1.6$\pm$1.1\% between ESPRESSO and HARPS, and $\sim$1.0$\pm$0.8\% between ESPRESSO and PEPSI). This makes the difference between PEPSI and HARPS EW measurements significant at a level of about 3$\sigma$.

We tested whether the observed differences in the line parameters between the spectrographs show a systematic dependence on wavelength or distance from the center of the echelle order (as measured in ESPRESSO 2D spectra). We found a small but statistically significant dependence of $\Delta$FWHM between ESPRESSO and HARPS on wavelength. This might be explained by a larger variation of the spectral resolution of ESPRESSO over the wavelength range than that of the HARPS spectrograph.

\begin{figure}
\begin{center}
\begin{tabular}{c}
\includegraphics[angle=0,width=0.9\linewidth]{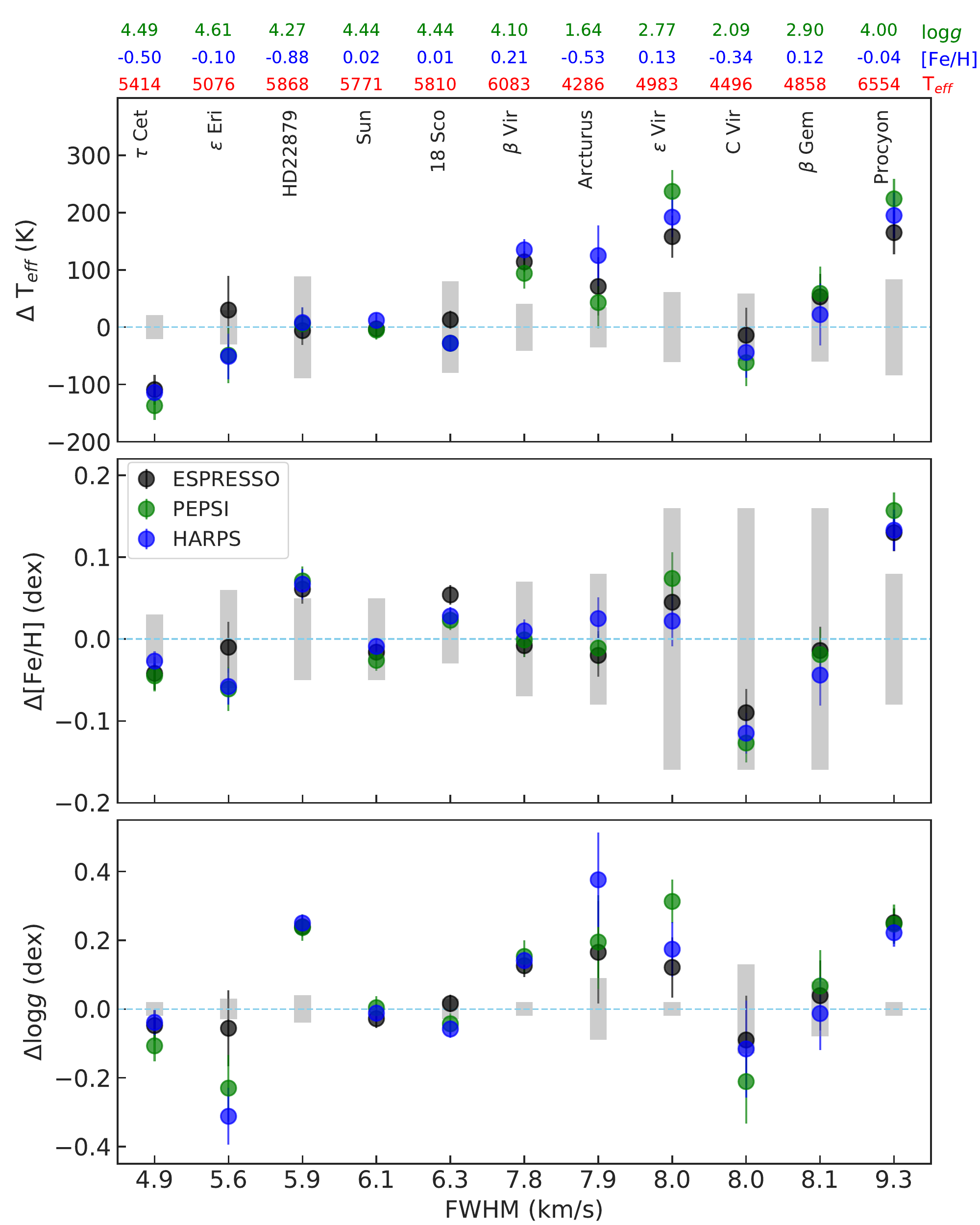}
\end{tabular}\end{center}
\vspace{-0.5cm}
\caption{Stellar atmospheric parameters relative to the reference values derived with the ESPRESSO (black), PEPSI (green), and HARPS (blue) spectra as a function of the mean FWHM of the lines. The atmospheric parameters of the stars are shown at the top of the panels. The reference values are from \citet{Jofre-14} and \citet{Heiter-15}. The gray areas represent the uncertainties of the reference parameters.}
\label{fig-param_comparision_all_stars}
\end{figure}

As in the previous section, we used these EW measurements to determine the stellar parameters of the sample stars and estimate the effect of the observed differences. Fig.~\ref{fig-param_comparision_all_stars} shows the stellar parameters of the stars as a function of the average FWHM of the spectral lines. The presented parameters are relative to the reference values as presented in \citet{Jofre-14} and \citet{Heiter-15}. Except for the significant differences from the reference values, our parameters derived from the ESPRESSO, PEPSI, and HARPS spectra agree well. The largest difference in stellar parameters observed between the spectrographs is as follows: $\Delta$\teff \ of 82$\pm$70 K for Arcturus (between PEPSI and HARPS); $\Delta$\logg \ of 0.256$\pm$0.138 dex for $\epsilon$ Eri (between ESPRESSO and HARPS); and $\Delta$[Fe/H] of 0.052$\pm$0.044 for  $\epsilon$ Vir (between PEPSI and HARPS). In general, the spectroscopic results obtained with ESPRESSO and PEPSI agree slightly better than with those obtained with HARPS.

It is beyond the scope of this manuscript to discuss the possible reasons for the observed deviation of these spectroscopic stellar parameters from the benchmark reference values in detail, and we refer to \citet{Jofre-14} and \citet{Heiter-15}. However, we recall that the reference values for \teff \ and \logg \ are not spectroscopic, but obtained with other  fundamental, less model-dependent methods. The only star for which our spectroscopic [Fe/H] deviates from the (spectroscopic) reference value by more than one sigma is Procyon. However, this is expected because of the observed difference in \teff \ and \logg \ for this star.

\section{Summary and conclusion}                                        \label{sec:summary}

We performed an intensive observational campaign with the main goal of creating a homogeneous UHR spectral library for all the Gaia benchmark stars and a catalog of homogeneously derived stellar parameters and chemical abundances of these stars. We obtained UHR ESPRESSO spectra for 30 out of the 34 FGK benchmark stars. The remaining 4 stars are not observable from the Paranal observatory, where the ESPRESSO spectrograph is installed. Fortunately, UHR spectra for these 4 stars (in addition to 19 more benchmark stars) have been obtained with the PEPSI spectrograph. Now the question that follows is ``Can PEPSI and ESPRESSO go well together?'' If for most of human beings it is just a matter of \textit{taste}, for us it is a matter of \textit{testing}. 

We compared the spectroscopic data obtained with ESPRESSO, PEPSI, and HARPS spectrographs. ESPRESSO and PESPI observations have been conducted at a spectral resolution of about $\sim$220\,000, and the resolution of HARPS spectra is R$\sim$115\,000. For our analysis we considered 11 Gaia benchmark stars with 4200 $<$ \teff \ $<$ 6600 K, 1.5 $<$ \logg \ $<$ 4.7 dex, -0.9 $<$ [Fe/H] $<$ 0.25 dex, and v$\sin i$ $<$ 10 km/s. The spectral lines (line depth, line width, and EW) were measured with the ARES v2 \citep{Sousa-15} automatic code.

In the first part of the manuscript we used the Vesta-reflected five solar spectra obtained with ESPRESSO to estimate the effect of  using different reduction products on the measurements of the spectral lines. Bellow we briefly present the main results in itemized form.

\begin{itemize}
 \item \textbf{Spectral lines observed in multiple echelle orders.} Our results suggest that on average, the EWs of spectral lines measured from different spectral orders agree well. However, we found an offset between the FWHM and depths of the spectral lines as measured in adjacent orders. We interpret this as a consequence of the variation in spectral resolution with wavelength for each echelle order. 
  \item \textbf{Spectral lines measured from the S1D and S2D spectra.} By measuring the EWs of spectral lines from the 1D and 2D spectra, we found an average difference of $\sim$0.6\% with a scatter that is slightly larger than the offset. The observed offset is larger for spectral lines that were observed in multiple orders (these lines are usually located farther from the center of the spectral order) than for lines that are located in only a single order.  Our comparison of these measurements with the  manual EW measurements of \citet{Sousa-08} based on the solar Kurucz atlas suggests a better agreement for the 1D spectra. This result may mean that a single measurement of spectral lines from combined (after combining the echelle orders and two slices) 1D spectra is preferable to simply averaging multiple measurements of spectral lines from different orders and slices in 2D spectra. 
  \item \textbf{Spectrum-to-spectrum scatter.} We compared the line parameters measured from five individual 1D spectra and found that the measurements agreed very well. The average scatter in EW is 0.35$\pm$0.15 m\AA,{} which is very close to the average statistical  photometric error for EWs estimated \citep{Cayrel-88} for the quality of the used spectra.
  \item \textbf{Solar atmospheric parameters and chemical abundances.} Using the EW  measurements of the lines from the five S1D spectra, we determined the stellar parameters and chemical abundances of a few refractory elements. Our results show an excellent agreement between the parameter and abundance determinations. A scatter of only 9 K for \teff, 0.01 dex for \logg, and 0.005 dex for [Fe/H] is measured. The observed scatter of the chemical abundances ranged from 0.002 dex (for \ion{Ti}{ii}) to 0.01 dex (for \ion{Al}{i} and \ion{Ca}{i}). The observed spectrum-to-spectrum abundance scatter for elements with only a few lines (equal to or fewer than three) is significantly smaller than the uncertainty based on the propagation of the EW errors, which takes into account the statistical photometric error and the error related to the continuum placement \citep{Cayrel-88}.
 \item \textbf{Individual versus combined spectra.} Our comparison of the solar atmospheric parameters and chemical abundances determined from the combined and individual S1D ESPRESSO spectra showed very consistent results. We also note that the uncertainties in the parameters and abundances (for the elements with several available spectral lines) do not decrease for the combined ESPRESSO spectrum when compared with those based on the individual, lower S/N ESPRESSO spectra. This result suggests that below given values, the precision is mostly dominated by the uncertainties in the atomic data of spectral lines, uncertainties in the model of atmospheres, our assumption of a Gaussian profile for the spectral lines, and/or the process of combining the individual spectra. However, it is important to note that for elements with only one or two weak lines, the increase in S/N is very important \citep[see, e.g.,][for the case of oxygen lines]{Adibekyan-16}. 
  
 \end{itemize}

In the second part of the paper we compared the spectroscopic results based on the measurements performed on the 1D combined ESPRESSO, PEPSI, and HARPS spectra. Below we itemized the main results.

\begin{itemize}
 \item \textbf{ESPRESSO -- PEPSI -- HARPS: spectral line measurements.} The comparison of the EW measurements based on the spectra of these three spectrographs showed a good agreement. The average relative difference in EW varied from 0.03$\pm$0.5\% (for $\beta$ Gem, between ESPRESSO and PEPSI) to 1.7$\pm$1.7\% (for HD22879 between ESPRESSO and PEPSI). While the EWs seems to be  conserved when line parameter measurements are made, our results show that the lines appear deeper in ESPRESSO spectra than in PEPSI and HARPS. The largest difference in line depth between ESPRESSO and PEPSI is about 3\% and about 9\% when ESPRESSO is compared with HARPS. The largest differences, as expected, are observed for the stars with intrinsically narrow spectral lines where the effect of the spectral resolution is more apparent. While the observed difference between ESPRESSO and HARPS is expected because of the difference in spectral resolutions, the discrepancy between ESPRESSO and PEPSI results is still to be understood.
 \item \textbf{ESPRESSO -- PEPSI -- HARPS: stellar parameters.} The good agreement in EW measurements between the different spectra resulted in consistent stellar parameters. The differences range from 2 to 82 K  in \teff, from 0.002 to 0.256 dex in \logg, and from 0.004 to 0.052 dex in [Fe/H].
 \end{itemize}

 Summarizing our results, we can conclude that the ESPRESSO, PEPSI, and HARPS spectrographs can deliver  spectroscopic results that are  sufficiently consistent for most of the sciences cases in stellar spectroscopy. However, we found small differences in the performance of the three spectrographs that can be important for specific science cases, for instance, when ultrahigh precision in chemical abundances based on limited weak lines is required, or when accurate line profile fitting is required. At this stage, a proper accounting for 3D effects and deviations from local thermodynamic equilibrium will be crucial to obtain reliable and accurate abudnances \citep[e.g.,][]{Bergemann-12, Amarsi-16}. These will be the subject of our next works based on the full ESPRESSO sample.

\begin{acknowledgements}
We thank the referee, Klaus G. Strassmeier, for the positive report and for providing extra (internal) information about the performance of the PEPSI spectrograph. This work was supported by FCT - Funda\c{c}\~ao para a Ci\^encia e Tecnologia (FCT) through national funds and by FEDER through COMPETE2020 - Programa Operacional Competitividade e Internacionaliza\c{c}\~ao by these grants: UID/FIS/04434/2019; UIDB/04434/2020; UIDP/04434/2020; PTDC/FIS-AST/32113/2017 \& POCI-01-0145-FEDER-032113; PTDC/FIS-AST/28953/2017 \& POCI-01-0145-FEDER-028953. V.A., E.D.M, N.C.S., and S.G.S. also acknowledge the support from FCT through Investigador FCT contracts nr.  IF/00650/2015/CP1273/CT0001, IF/00849/2015/CP1273/CT0003, IF/00169/2012/CP0150/CT0002,   and IF/00028/2014/CP1215/CT0002, respectively, and POPH/FSE (EC) by FEDER funding through the program ``Programa Operacional de Factores de Competitividade - COMPETE''. TLC acknowledges support from the European Union's Horizon 2020 research and innovation programme under the Marie Sk\l{}odowska-Curie grant agreement No.~792848 (PULSATION). CAP and JIGH are thankful to the Spanish government for research funding (AYA2017-86389-P).
\end{acknowledgements}

\bibliographystyle{aa}
\bibliography{references}

\begin{thebibliography}{66}
\expandafter\ifx\csname natexlab\endcsname\relax\def\natexlab#1{#1}\fi

\bibitem[{{Adibekyan} {et~al.}(2018){Adibekyan}, {de Laverny}, {Recio-Blanco},
  {Sousa}, {Delgado-Mena}, {Kordopatis}, {Ferreira}, {Santos}, {Hakobyan}, \&
  {Tsantaki}}]{Adibekyan-18}
{Adibekyan}, V., {de Laverny}, P., {Recio-Blanco}, A., {et~al.} 2018, \aap,
  619, A130

\bibitem[{{Adibekyan} {et~al.}(2016){Adibekyan}, {Delgado-Mena}, {Figueira},
  {Sousa}, {Santos}, {Faria}, {Gonz{\'a}lez Hern{\'a}ndez}, {Israelian},
  {Harutyunyan}, {Su{\'a}rez-Andr{\'e}s}, \& {Hakobyan}}]{Adibekyan-16}
{Adibekyan}, V., {Delgado-Mena}, E., {Figueira}, P., {et~al.} 2016, \aap, 591,
  A34

\bibitem[{{Adibekyan} {et~al.}(2015{\natexlab{a}}){Adibekyan}, {Figueira},
  {Santos}, {Sousa}, {Faria}, {Delgado-Mena}, {Oshagh}, {Tsantaki}, {Hakobyan},
  {Gonz{\'a}lez Hern{\'a}ndez}, {Su{\'a}rez-Andr{\'e}s}, \&
  {Israelian}}]{Adibekyan-15a}
{Adibekyan}, V., {Figueira}, P., {Santos}, N.~C., {et~al.} 2015{\natexlab{a}},
  \aap, 583, A94

\bibitem[{{Adibekyan} {et~al.}(2015{\natexlab{b}}){Adibekyan}, {Santos},
  {Figueira}, {Dorn}, {Sousa}, {Delgado-Mena}, {Israelian}, {Hakobyan}, \&
  {Mordasini}}]{Adibekyan-15}
{Adibekyan}, V., {Santos}, N.~C., {Figueira}, P., {et~al.} 2015{\natexlab{b}},
  \aap, 581, L2

\bibitem[{{Adibekyan} {et~al.}(2012{\natexlab{a}}){Adibekyan}, {Delgado Mena},
  {Sousa}, {Santos}, {Israelian}, {Gonz{\'a}lez Hern{\'a}ndez}, {Mayor}, \&
  {Hakobyan}}]{Adibekyan-12_kepler}
{Adibekyan}, V.~Z., {Delgado Mena}, E., {Sousa}, S.~G., {et~al.}
  2012{\natexlab{a}}, \aap, 547, A36

\bibitem[{{Adibekyan} {et~al.}(2013){Adibekyan}, {Figueira}, {Santos},
  {Hakobyan}, {Sousa}, {Pace}, {Delgado Mena}, {Robin}, {Israelian}, \&
  {Gonz{\'a}lez Hern{\'a}ndez}}]{Adibekyan-13}
{Adibekyan}, V.~Z., {Figueira}, P., {Santos}, N.~C., {et~al.} 2013, \aap, 554,
  A44

\bibitem[{{Adibekyan} {et~al.}(2011){Adibekyan}, {Santos}, {Sousa}, \&
  {Israelian}}]{Adibekyan-11}
{Adibekyan}, V.~Z., {Santos}, N.~C., {Sousa}, S.~G., \& {Israelian}, G. 2011,
  \aap, 535, L11

\bibitem[{{Adibekyan} {et~al.}(2012{\natexlab{b}}){Adibekyan}, {Santos},
  {Sousa}, {Israelian}, {Delgado Mena}, {Gonz{\'a}lez Hern{\'a}ndez}, {Mayor},
  {Lovis}, \& {Udry}}]{Adibekyan-12_harps}
{Adibekyan}, V.~Z., {Santos}, N.~C., {Sousa}, S.~G., {et~al.}
  2012{\natexlab{b}}, \aap, 543, A89

\bibitem[{{Adibekyan} {et~al.}(2012{\natexlab{c}}){Adibekyan}, {Sousa},
  {Santos}, {Delgado Mena}, {Gonz{\'a}lez Hern{\'a}ndez}, {Israelian}, {Mayor},
  \& {Khachatryan}}]{Adibekyan-12_1111}
{Adibekyan}, V.~Z., {Sousa}, S.~G., {Santos}, N.~C., {et~al.}
  2012{\natexlab{c}}, \aap, 545, A32

\bibitem[{{Amarsi} {et~al.}(2016){Amarsi}, {Lind}, {Asplund}, {Barklem}, \&
  {Collet}}]{Amarsi-16}
{Amarsi}, A.~M., {Lind}, K., {Asplund}, M., {Barklem}, P.~S., \& {Collet}, R.
  2016, \mnras, 463, 1518

\bibitem[{{Bedell} {et~al.}(2014){Bedell}, {Mel{\'e}ndez}, {Bean},
  {Ram{\'\i}rez}, {Leite}, \& {Asplund}}]{Bedell-14}
{Bedell}, M., {Mel{\'e}ndez}, J., {Bean}, J.~L., {et~al.} 2014, \apj, 795, 23

\bibitem[{{Bensby} {et~al.}(2014){Bensby}, {Feltzing}, \& {Oey}}]{Bensby-14}
{Bensby}, T., {Feltzing}, S., \& {Oey}, M.~S. 2014, \aap, 562, A71

\bibitem[{{Bergemann} {et~al.}(2012){Bergemann}, {Lind}, {Collet}, {Magic}, \&
  {Asplund}}]{Bergemann-12}
{Bergemann}, M., {Lind}, K., {Collet}, R., {Magic}, Z., \& {Asplund}, M. 2012,
  \mnras, 427, 27

\bibitem[{{Bergemann} {et~al.}(2014){Bergemann}, {Ruchti}, {Serenelli},
  {Feltzing}, {Alves-Brito}, {Asplund}, {Bensby}, {Gruyters}, {Heiter},
  {Hourihane}, {Korn}, {Lind}, {Marino}, {Jofre}, {Nordland er}, {Ryde},
  {Worley}, {Gilmore}, {Rand ich}, {Ferguson}, {Jeffries}, {Micela},
  {Negueruela}, {Prusti}, {Rix}, {Vallenari}, {Alfaro}, {Allende Prieto},
  {Bragaglia}, {Koposov}, {Lanzafame}, {Pancino}, {Recio-Blanco}, {Smiljanic},
  {Walton}, {Costado}, {Franciosini}, {Hill}, {Lardo}, {de Laverny}, {Magrini},
  {Maiorca}, {Masseron}, {Morbidelli}, {Sacco}, {Kordopatis}, \&
  {Tautvai{\v{s}}ien{\.{e}}}}]{Bergemann-14}
{Bergemann}, M., {Ruchti}, G.~R., {Serenelli}, A., {et~al.} 2014, \aap, 565,
  A89

\bibitem[{{Bertran de Lis} {et~al.}(2015){Bertran de Lis}, {Delgado Mena},
  {Adibekyan}, {Santos}, \& {Sousa}}]{BertrandeLis-15}
{Bertran de Lis}, S., {Delgado Mena}, E., {Adibekyan}, V.~Z., {Santos}, N.~C.,
  \& {Sousa}, S.~G. 2015, \aap, 576, A89

\bibitem[{{Blanco-Cuaresma} {et~al.}(2014){Blanco-Cuaresma}, {Soubiran},
  {Jofr{\'e}}, \& {Heiter}}]{Blanco-Cuaresma-14}
{Blanco-Cuaresma}, S., {Soubiran}, C., {Jofr{\'e}}, P., \& {Heiter}, U. 2014,
  \aap, 566, A98

\bibitem[{{Brahm} {et~al.}(2017){Brahm}, {Jord{\'a}n}, \&
  {Espinoza}}]{Brahm-17}
{Brahm}, R., {Jord{\'a}n}, A., \& {Espinoza}, N. 2017, \pasp, 129, 034002

\bibitem[{{Cayrel}(1988)}]{Cayrel-88}
{Cayrel}, R. 1988, in IAU Symposium, Vol. 132, The Impact of Very High S/N
  Spectroscopy on Stellar Physics, ed. G.~{Cayrel de Strobel} \& M.~{Spite},
  345

\bibitem[{{de Laverny} {et~al.}(2012){de Laverny}, {Recio-Blanco}, {Worley}, \&
  {Plez}}]{deLaverny-12}
{de Laverny}, P., {Recio-Blanco}, A., {Worley}, C.~C., \& {Plez}, B. 2012,
  \aap, 544, A126

\bibitem[{{De Silva} {et~al.}(2015){De Silva}, {Freeman}, {Bland-Hawthorn},
  {Martell}, {de Boer}, {Asplund}, {Keller}, {Sharma}, {Zucker}, {Zwitter},
  {Anguiano}, {Bacigalupo}, {Bayliss}, {Beavis}, {Bergemann}, {Campbell},
  {Cannon}, {Carollo}, {Casagrande}, {Casey}, {Da Costa}, {D'Orazi}, {Dotter},
  {Duong}, {Heger}, {Ireland}, {Kafle}, {Kos}, {Lattanzio}, {Lewis}, {Lin},
  {Lind}, {Munari}, {Nataf}, {O'Toole}, {Parker}, {Reid}, {Schlesinger},
  {Sheinis}, {Simpson}, {Stello}, {Ting}, {Traven}, {Watson}, {Wittenmyer},
  {Yong}, \& {{\v{Z}}erjal}}]{DeSilva-15}
{De Silva}, G.~M., {Freeman}, K.~C., {Bland-Hawthorn}, J., {et~al.} 2015,
  \mnras, 449, 2604

\bibitem[{{Delgado Mena} {et~al.}(2018){Delgado Mena}, {Adibekyan}, {Figueira},
  {Gonz{\'a}lez Hern{\'a}ndez}, {Santos}, {Tsantaki}, {Sousa}, {Faria},
  {Su{\'a}rez-Andr{\'e}s}, \& {Israelian}}]{Delgado-Mena-18}
{Delgado Mena}, E., {Adibekyan}, V.~Z., {Figueira}, P., {et~al.} 2018, \pasp,
  130, 094202

\bibitem[{{Delgado Mena} {et~al.}(2017){Delgado Mena}, {Tsantaki}, {Adibekyan},
  {Sousa}, {Santos}, {Gonz{\'a}lez Hern{\'a}ndez}, \&
  {Israelian}}]{Delgado-Mena-17}
{Delgado Mena}, E., {Tsantaki}, M., {Adibekyan}, V.~Z., {et~al.} 2017, \aap,
  606, A94

\bibitem[{{Deng} {et~al.}(2012){Deng}, {Newberg}, {Liu}, {Carlin}, {Beers},
  {Chen}, {Chen}, {Christlieb}, {Grillmair}, {Guhathakurta}, {Han}, {Hou},
  {Lee}, {L{\'e}pine}, {Li}, {Liu}, {Pan}, {Sellwood}, {Wang}, {Wang}, {Yang},
  {Yanny}, {Zhang}, {Zhang}, {Zheng}, \& {Zhu}}]{Deng-12}
{Deng}, L.-C., {Newberg}, H.~J., {Liu}, C., {et~al.} 2012, Research in
  Astronomy and Astrophysics, 12, 735

\bibitem[{{Erspamer} \& {North}(2002)}]{Erspamer-02}
{Erspamer}, D. \& {North}, P. 2002, \aap, 383, 227

\bibitem[{Eversberg \& Vollmann(2015)}]{Eversberg2015}
Eversberg, T. \& Vollmann, K. 2015, Fundamentals of Echelle Spectroscopy
  (Berlin, Heidelberg: Springer Berlin Heidelberg), 193--227

\bibitem[{{Gilmore} {et~al.}(2012){Gilmore}, {Randich}, {Asplund}, {Binney},
  {Bonifacio}, {Drew}, {Feltzing}, {Ferguson}, {Jeffries}, {Micela},
  {Negueruela}, {Prusti}, {Rix}, {Vallenari}, {Alfaro}, {Allende-Prieto},
  {Babusiaux}, {Bensby}, {Blomme}, {Bragaglia}, {Flaccomio}, {Fran{\c{c}}ois},
  {Irwin}, {Koposov}, {Korn}, {Lanzafame}, {Pancino}, {Paunzen},
  {Recio-Blanco}, {Sacco}, {Smiljanic}, {Van Eck}, {Walton}, {Aden}, {Aerts},
  {Affer}, {Alcala}, {Altavilla}, {Alves}, {Antoja}, {Arenou}, {Argiroffi},
  {Asensio Ramos}, {Bailer-Jones}, {Balaguer-Nunez}, {Bayo}, {Barbuy},
  {Barisevicius}, {Barrado y Navascues}, {Battistini}, {Bellas Velidis},
  {Bellazzini}, {Belokurov}, {Bergemann}, {Bertelli}, {Biazzo}, {Bienayme},
  {Bland-Hawthorn}, {Boeche}, {Bonito}, {Boudreault}, {Bouvier}, {Brandao},
  {Brown}, {de Bruijne}, {Burleigh}, {Caballero}, {Caffau}, {Calura},
  {Capuzzo-Dolcetta}, {Caramazza}, {Carraro}, {Casagrande}, {Casewell},
  {Chapman}, {Chiappini}, {Chorniy}, {Christlieb}, {Cignoni}, {Cocozza},
  {Colless}, {Collet}, {Collins}, {Correnti}, {Covino}, {Crnojevic}, {Cropper},
  {Cunha}, {Damiani}, {David}, {Delgado}, {Duffau}, {Edvardsson}, {Eldridge},
  {Enke}, {Eriksson}, {Evans}, {Eyer}, {Famaey}, {Fellhauer}, {Ferreras},
  {Figueras}, {Fiorentino}, {Flynn}, {Folha}, {Franciosini}, {Frasca},
  {Freeman}, {Fremat}, {Friel}, {Gaensicke}, {Gameiro}, {Garzon}, {Geier},
  {Geisler}, {Gerhard}, {Gibson}, {Gomboc}, {Gomez}, {Gonzalez-Fernandez},
  {Gonzalez Hernandez}, {Gosset}, {Grebel}, {Greimel}, {Groenewegen},
  {Grundahl}, {Guarcello}, {Gustafsson}, {Hadrava}, {Hatzidimitriou}, {Hambly},
  {Hammersley}, {Hansen}, {Haywood}, {Heber}, {Heiter}, {Held}, {Helmi},
  {Hensler}, {Herrero}, {Hill}, {Hodgkin}, {Huelamo}, {Huxor}, {Ibata},
  {Jackson}, {de Jong}, {Jonker}, {Jordan}, {Jordi}, {Jorissen}, {Katz},
  {Kawata}, {Keller}, {Kharchenko}, {Klement}, {Klutsch}, {Knude}, {Koch},
  {Kochukhov}, {Kontizas}, {Koubsky}, {Lallement}, {de Laverny}, {van Leeuwen},
  {Lemasle}, {Lewis}, {Lind}, {Lindstrom}, {Lobel}, {Lopez Santiago}, {Lucas},
  {Ludwig}, {Lueftinger}, {Magrini}, {Maiz Apellaniz}, {Maldonado}, {Marconi},
  {Marino}, {Martayan}, {Martinez-Valpuesta}, {Matijevic}, {McMahon},
  {Messina}, {Meyer}, {Miglio}, {Mikolaitis}, {Minchev}, {Minniti}, {Moitinho},
  {Momany}, {Monaco}, {Montalto}, {Monteiro}, {Monier}, {Montes}, {Mora},
  {Moraux}, {Morel}, {Mowlavi}, {Mucciarelli}, {Munari}, {Napiwotzki},
  {Nardetto}, {Naylor}, {Naze}, {Nelemans}, {Okamoto}, {Ortolani}, {Pace},
  {Palla}, {Palous}, {Parker}, {Penarrubia}, {Pillitteri}, {Piotto}, {Posbic},
  {Prisinzano}, {Puzeras}, {Quirrenbach}, {Ragaini}, {Read}, {Read}, {Reyle},
  {De Ridder}, {Robichon}, {Robin}, {Roeser}, {Romano}, {Royer}, {Ruchti},
  {Ruzicka}, {Ryan}, {Ryde}, {Santos}, {Sanz Forcada}, {Sarro Baro},
  {Sbordone}, {Schilbach}, {Schmeja}, {Schnurr}, {Schoenrich}, {Scholz},
  {Seabroke}, {Sharma}, {De Silva}, {Smith}, {Solano}, {Sordo}, {Soubiran},
  {Sousa}, {Spagna}, {Steffen}, {Steinmetz}, {Stelzer}, {Stempels},
  {Tabernero}, {Tautvaisiene}, {Thevenin}, {Torra}, {Tosi}, {Tolstoy}, {Turon},
  {Walker}, {Wambsganss}, {Worley}, {Venn}, {Vink}, {Wyse}, {Zaggia},
  {Zeilinger}, {Zoccali}, {Zorec}, {Zucker}, {Zwitter}, \& {Gaia-ESO Survey
  Team}}]{Gilmore-12}
{Gilmore}, G., {Randich}, S., {Asplund}, M., {et~al.} 2012, The Messenger, 147,
  25

\bibitem[{{Gonz{\'a}lez Hern{\'a}ndez} {et~al.}(2013){Gonz{\'a}lez
  Hern{\'a}ndez}, {Delgado-Mena}, {Sousa}, {Israelian}, {Santos}, {Adibekyan},
  \& {Udry}}]{Gonzalez-Hernandez-13}
{Gonz{\'a}lez Hern{\'a}ndez}, J.~I., {Delgado-Mena}, E., {Sousa}, S.~G.,
  {et~al.} 2013, \aap, 552, A6

\bibitem[{{Gonz{\'a}lez Hern{\'a}ndez} {et~al.}(2010){Gonz{\'a}lez
  Hern{\'a}ndez}, {Israelian}, {Santos}, {Sousa}, {Delgado-Mena}, {Neves}, \&
  {Udry}}]{Gonzalez-Hernandez-10}
{Gonz{\'a}lez Hern{\'a}ndez}, J.~I., {Israelian}, G., {Santos}, N.~C., {et~al.}
  2010, \apj, 720, 1592

\bibitem[{{Gonz{\'a}lez Hern{\'a}ndez} {et~al.}(2018){Gonz{\'a}lez
  Hern{\'a}ndez}, {Pepe}, {Molaro}, \& {Santos}}]{Gonzalez-Hernandez-18}
{Gonz{\'a}lez Hern{\'a}ndez}, J.~I., {Pepe}, F., {Molaro}, P., \& {Santos},
  N.~C. 2018, {ESPRESSO on VLT: An Instrument for Exoplanet Research, Handbook
  of Exoplanets, ISBN 978-3-319-55332-0. Springer International Publishing AG,
  part of Springer Nature, 2018, id.157}, 157

\bibitem[{{Hawkins} {et~al.}(2016){Hawkins}, {Jofr{\'e}}, {Heiter}, {Soubiran},
  {Blanco-Cuaresma}, {Casagrande}, {Gilmore}, {Lind}, {Magrini}, {Masseron},
  {Pancino}, {Randich}, \& {Worley}}]{Hawkins-16}
{Hawkins}, K., {Jofr{\'e}}, P., {Heiter}, U., {et~al.} 2016, \aap, 592, A70

\bibitem[{{Heiter} {et~al.}(2015){Heiter}, {Jofr{\'e}}, {Gustafsson}, {Korn},
  {Soubiran}, \& {Th{\'e}venin}}]{Heiter-15}
{Heiter}, U., {Jofr{\'e}}, P., {Gustafsson}, B., {et~al.} 2015, \aap, 582, A49

\bibitem[{{Hensberge}(2004)}]{Hensberge-04}
{Hensberge}, H. 2004, in Astronomical Society of the Pacific Conference Series,
  Vol. 318, Spectroscopically and Spatially Resolving the Components of the
  Close Binary Stars, ed. R.~W. {Hilditch}, H.~{Hensberge}, \& K.~{Pavlovski},
  43--51

\bibitem[{{Hensberge}(2007)}]{Hensberge-07}
{Hensberge}, H. 2007, in Astronomical Society of the Pacific Conference Series,
  Vol. 364, The Future of Photometric, Spectrophotometric and Polarimetric
  Standardization, ed. C.~{Sterken}, 275

\bibitem[{{Hinkle} {et~al.}(2000){Hinkle}, {Wallace}, {Valenti}, \&
  {Harmer}}]{Hinkle-00}
{Hinkle}, K., {Wallace}, L., {Valenti}, J., \& {Harmer}, D. 2000, {Visible and
  Near Infrared Atlas of the Arcturus Spectrum 3727-9300 A ed. Kenneth Hinkle,
  Lloyd Wallace, Jeff Valenti, and Dianne Harmer. (San Francisco: ASP) ISBN:
  1-58381-037-4, 2000}

\bibitem[{{Jofr{\'e}} {et~al.}(2015){Jofr{\'e}}, {Heiter}, {Soubiran},
  {Blanco-Cuaresma}, {Masseron}, {Nordlander}, {Chemin}, {Worley}, {Van Eck},
  {Hourihane}, {Gilmore}, {Adibekyan}, {Bergemann}, {Cantat-Gaudin},
  {Delgado-Mena}, {Gonz{\'a}lez Hern{\'a}ndez}, {Guiglion}, {Lardo}, {de
  Laverny}, {Lind}, {Magrini}, {Mikolaitis}, {Montes}, {Pancino},
  {Recio-Blanco}, {Sordo}, {Sousa}, {Tabernero}, \& {Vallenari}}]{Jofre-15}
{Jofr{\'e}}, P., {Heiter}, U., {Soubiran}, C., {et~al.} 2015, \aap, 582, A81

\bibitem[{{Jofr{\'e}} {et~al.}(2014){Jofr{\'e}}, {Heiter}, {Soubiran},
  {Blanco-Cuaresma}, {Worley}, {Pancino}, {Cantat-Gaudin}, {Magrini},
  {Bergemann}, {Gonz{\'a}lez Hern{\'a}ndez}, {Hill}, {Lardo}, {de Laverny},
  {Lind}, {Masseron}, {Montes}, {Mucciarelli}, {Nordlander}, {Recio Blanco},
  {Sobeck}, {Sordo}, {Sousa}, {Tabernero}, {Vallenari}, \& {Van
  Eck}}]{Jofre-14}
{Jofr{\'e}}, P., {Heiter}, U., {Soubiran}, C., {et~al.} 2014, \aap, 564, A133

\bibitem[{{Jofr{\'e}} {et~al.}(2017){Jofr{\'e}}, {Heiter}, {Worley},
  {Blanco-Cuaresma}, {Soubiran}, {Masseron}, {Hawkins}, {Adibekyan}, {Buder},
  {Casamiquela}, {Gilmore}, {Hourihane}, \& {Tabernero}}]{Jofre-17}
{Jofr{\'e}}, P., {Heiter}, U., {Worley}, C.~C., {et~al.} 2017, \aap, 601, A38

\bibitem[{{Kurucz}(1993)}]{Kurucz-93}
{Kurucz}, R. 1993, ATLAS9 Stellar Atmosphere Programs and 2 km/s grid. Kurucz
  CD-ROM No. 13. Cambridge, 13

\bibitem[{{Kurucz}(2005)}]{Kurucz-05}
{Kurucz}, R.~L. 2005, Memorie della Societa Astronomica Italiana Supplementi,
  8, 189

\bibitem[{{Liu} {et~al.}(2020){Liu}, {Yong}, {Asplund}, {Wang}, {Spina},
  {Acu{\~n}a}, {Mel{\'e}ndez}, \& {Ram{\'\i}rez}}]{Liu-20}
{Liu}, F., {Yong}, D., {Asplund}, M., {et~al.} 2020, \mnras

\bibitem[{{Majewski} {et~al.}(2017){Majewski}, {Schiavon}, {Frinchaboy},
  {Allende Prieto}, {Barkhouser}, {Bizyaev}, {Blank}, {Brunner}, {Burton},
  {Carrera}, {Chojnowski}, {Cunha}, {Epstein}, {Fitzgerald}, {Garc{\'\i}a
  P{\'e}rez}, {Hearty}, {Henderson}, {Holtzman}, {Johnson}, {Lam}, {Lawler},
  {Maseman}, {M{\'e}sz{\'a}ros}, {Nelson}, {Nguyen}, {Nidever}, {Pinsonneault},
  {Shetrone}, {Smee}, {Smith}, {Stolberg}, {Skrutskie}, {Walker}, {Wilson},
  {Zasowski}, {Anders}, {Basu}, {Beland}, {Blanton}, {Bovy}, {Brownstein},
  {Carlberg}, {Chaplin}, {Chiappini}, {Eisenstein}, {Elsworth}, {Feuillet},
  {Fleming}, {Galbraith-Frew}, {Garc{\'\i}a}, {Garc{\'\i}a-Hern{\'a}ndez},
  {Gillespie}, {Girardi}, {Gunn}, {Hasselquist}, {Hayden}, {Hekker}, {Ivans},
  {Kinemuchi}, {Klaene}, {Mahadevan}, {Mathur}, {Mosser}, {Muna}, {Munn},
  {Nichol}, {O'Connell}, {Parejko}, {Robin}, {Rocha-Pinto}, {Schultheis},
  {Serenelli}, {Shane}, {Silva Aguirre}, {Sobeck}, {Thompson}, {Troup},
  {Weinberg}, \& {Zamora}}]{Majewski-17}
{Majewski}, S.~R., {Schiavon}, R.~P., {Frinchaboy}, P.~M., {et~al.} 2017, \aj,
  154, 94

\bibitem[{{Mayor} {et~al.}(2003){Mayor}, {Pepe}, {Queloz}, {Bouchy},
  {Rupprecht}, {Lo Curto}, {Avila}, {Benz}, {Bertaux}, {Bonfils}, {Dall},
  {Dekker}, {Delabre}, {Eckert}, {Fleury}, {Gilliotte}, {Gojak}, {Guzman},
  {Kohler}, {Lizon}, {Longinotti}, {Lovis}, {Megevand}, {Pasquini}, {Reyes},
  {Sivan}, {Sosnowska}, {Soto}, {Udry}, {van Kesteren}, {Weber}, \&
  {Weilenmann}}]{Mayor-03}
{Mayor}, M., {Pepe}, F., {Queloz}, D., {et~al.} 2003, The Messenger, 114, 20

\bibitem[{{Nissen}(2015)}]{Nissen-15}
{Nissen}, P.~E. 2015, \aap, 579, A52

\bibitem[{{Nissen} {et~al.}(2020){Nissen}, {Christensen-Dalsgaard},
  {Mosumgaard}, {Silva Aguirre}, {Spitoni}, \& {Verma.}}]{Nissen-20}
{Nissen}, P.~E., {Christensen-Dalsgaard}, J., {Mosumgaard}, J.~R., {et~al.}
  2020, arXiv e-prints, arXiv:2006.06013

\bibitem[{{Nissen} \& {Schuster}(2010)}]{Nissen-10}
{Nissen}, P.~E. \& {Schuster}, W.~J. 2010, \aap, 511, L10

\bibitem[{{Pepe} {et~al.}(2010){Pepe}, {Cristiani}, {Rebolo Lopez}, {Santos},
  {Amorim}, {Avila}, {Benz}, {Bonifacio}, {Cabral}, {Carvas}, {Cirami},
  {Coelho}, {Comari}, {Coretti}, {De Caprio}, {Dekker}, {Delabre}, {Di
  Marcantonio}, {D'Odorico}, {Fleury}, {Garc{\'\i}a}, {Herreros Linares},
  {Hughes}, {Iwert}, {Lima}, {Lizon}, {Lo Curto}, {Lovis}, {Manescau},
  {Martins}, {M{\'e}gevand}, {Moitinho}, {Molaro}, {Monteiro}, {Monteiro},
  {Pasquini}, {Mordasini}, {Queloz}, {Rasilla}, {Rebord{\~a}o}, {Santana
  Tschudi}, {Santin}, {Sosnowska}, {Span{\`o}}, {Tenegi}, {Udry}, {Vanzella},
  {Viel}, {Zapatero Osorio}, \& {Zerbi}}]{Pepe-10}
{Pepe}, F.~A., {Cristiani}, S., {Rebolo Lopez}, R., {et~al.} 2010, in Society
  of Photo-Optical Instrumentation Engineers (SPIE) Conference Series, Vol.
  7735, \procspie, 77350F

\bibitem[{{Perryman} {et~al.}(2001){Perryman}, {de Boer}, {Gilmore}, {H{\o}g},
  {Lattanzi}, {Lindegren}, {Luri}, {Mignard}, {Pace}, \& {de
  Zeeuw}}]{Perryman-01}
{Perryman}, M.~A.~C., {de Boer}, K.~S., {Gilmore}, G., {et~al.} 2001, \aap,
  369, 339

\bibitem[{{Piskunov} \& {Valenti}(2002)}]{Piskunov-02}
{Piskunov}, N.~E. \& {Valenti}, J.~A. 2002, \aap, 385, 1095

\bibitem[{{Privitera} {et~al.}(2016){Privitera}, {Meynet}, {Eggenberger},
  {Vidotto}, {Villaver}, \& {Bianda}}]{Privitera-16}
{Privitera}, G., {Meynet}, G., {Eggenberger}, P., {et~al.} 2016, \aap, 593,
  A128

\bibitem[{{Prugniel} \& {Soubiran}(2001)}]{Prugniel-01}
{Prugniel}, P. \& {Soubiran}, C. 2001, \aap, 369, 1048

\bibitem[{{Ram{\'\i}rez} {et~al.}(2019){Ram{\'\i}rez}, {Khanal}, {Lichon},
  {Chanam{\'e}}, {Endl}, {Mel{\'e}ndez}, \& {Lambert}}]{Ramirez-19}
{Ram{\'\i}rez}, I., {Khanal}, S., {Lichon}, S.~J., {et~al.} 2019, \mnras, 490,
  2448

\bibitem[{{Saffe} {et~al.}(2016){Saffe}, {Flores}, {Jaque Arancibia},
  {Buccino}, \& {Jofr{\'e}}}]{Saffe-16}
{Saffe}, C., {Flores}, M., {Jaque Arancibia}, M., {Buccino}, A., \&
  {Jofr{\'e}}, E. 2016, \aap, 588, A81

\bibitem[{{Sneden}(1973)}]{Sneden-73}
{Sneden}, C.~A. 1973, PhD thesis, THE UNIVERSITY OF TEXAS AT AUSTIN.

\bibitem[{{Sousa}(2014)}]{Sousa-14}
{Sousa}, S.~G. 2014, {ARES + MOOG: A Practical Overview of an Equivalent Width
  (EW) Method to Derive Stellar Parameters}, Springer International Publishing
  (Cham), Edited by Ewa Niemczura, Barry Smalley and Wojtek Pych, pp. 297--310

\bibitem[{{Sousa} {et~al.}(2018){Sousa}, {Adibekyan}, {Delgado-Mena}, {Santos},
  {Andreasen}, {Ferreira}, {Tsantaki}, {Barros}, {Demangeon}, {Israelian},
  {Faria}, {Figueira}, {Mortier}, {Brand{\~a}o}, {Montalto}, {Rojas-Ayala}, \&
  {Santerne}}]{Sousa-18}
{Sousa}, S.~G., {Adibekyan}, V., {Delgado-Mena}, E., {et~al.} 2018, \aap, 620,
  A58

\bibitem[{{Sousa} {et~al.}(2015{\natexlab{a}}){Sousa}, {Santos}, {Adibekyan},
  {Delgado-Mena}, \& {Israelian}}]{Sousa-15}
{Sousa}, S.~G., {Santos}, N.~C., {Adibekyan}, V., {Delgado-Mena}, E., \&
  {Israelian}, G. 2015{\natexlab{a}}, \aap, 577, A67

\bibitem[{{Sousa} {et~al.}(2008){Sousa}, {Santos}, {Mayor}, {Udry},
  {Casagrande}, {Israelian}, {Pepe}, {Queloz}, \& {Monteiro}}]{Sousa-08}
{Sousa}, S.~G., {Santos}, N.~C., {Mayor}, M., {et~al.} 2008, \aap, 487, 373

\bibitem[{{Sousa} {et~al.}(2015{\natexlab{b}}){Sousa}, {Santos}, {Mortier},
  {Tsantaki}, {Adibekyan}, {Delgado Mena}, {Israelian}, {Rojas-Ayala}, \&
  {Neves}}]{Sousa-15a}
{Sousa}, S.~G., {Santos}, N.~C., {Mortier}, A., {et~al.} 2015{\natexlab{b}},
  \aap, 576, A94

\bibitem[{{Spina} {et~al.}(2016){Spina}, {Mel{\'e}ndez}, {Karakas},
  {Ram{\'\i}rez}, {Monroe}, {Asplund}, \& {Yong}}]{Spina-16}
{Spina}, L., {Mel{\'e}ndez}, J., {Karakas}, A.~I., {et~al.} 2016, \aap, 593,
  A125

\bibitem[{{Steinmetz} {et~al.}(2006){Steinmetz}, {Zwitter}, {Siebert},
  {Watson}, {Freeman}, {Munari}, {Campbell}, {Williams}, {Seabroke}, {Wyse},
  {Parker}, {Bienaym{\'e}}, {Roeser}, {Gibson}, {Gilmore}, {Grebel}, {Helmi},
  {Navarro}, {Burton}, {Cass}, {Dawe}, {Fiegert}, {Hartley}, {Russell},
  {Saunders}, {Enke}, {Bailin}, {Binney}, {Bland -Hawthorn}, {Boeche},
  {Dehnen}, {Eisenstein}, {Evans}, {Fiorucci}, {Fulbright}, {Gerhard},
  {Jauregi}, {Kelz}, {Mijovi{\'c}}, {Minchev}, {Parmentier}, {Pe{\~n}arrubia},
  {Quillen}, {Read}, {Ruchti}, {Scholz}, {Siviero}, {Smith}, {Sordo}, {Veltz},
  {Vidrih}, {von Berlepsch}, {Boyle}, \& {Schilbach}}]{Steinmetz-06}
{Steinmetz}, M., {Zwitter}, T., {Siebert}, A., {et~al.} 2006, \aj, 132, 1645

\bibitem[{{Strassmeier} {et~al.}(2015){Strassmeier}, {Ilyin}, {J{\"a}rvinen},
  {Weber}, {Woche}, {Barnes}, {Bauer}, {Beckert}, {Bittner}, {Bredthauer},
  {Carroll}, {Denker}, {Dionies}, {DiVarano}, {D{\"o}scher}, {Fechner},
  {Feuerstein}, {Granzer}, {Hahn}, {Harnisch}, {Hofmann}, {Lesser}, {Paschke},
  {Pankratow}, {Plank}, {Pl{\"u}schke}, {Popow}, \&
  {Sablowski}}]{Strassmeier-15AN}
{Strassmeier}, K.~G., {Ilyin}, I., {J{\"a}rvinen}, A., {et~al.} 2015,
  Astronomische Nachrichten, 336, 324

\bibitem[{{Strassmeier} {et~al.}(2018{\natexlab{a}}){Strassmeier}, {Ilyin}, \&
  {Steffen}}]{Strassmeier-18_sun}
{Strassmeier}, K.~G., {Ilyin}, I., \& {Steffen}, M. 2018{\natexlab{a}}, \aap,
  612, A44

\bibitem[{{Strassmeier} {et~al.}(2018{\natexlab{b}}){Strassmeier}, {Ilyin}, \&
  {Weber}}]{Strassmeier-18}
{Strassmeier}, K.~G., {Ilyin}, I., \& {Weber}, M. 2018{\natexlab{b}}, \aap,
  612, A45

\bibitem[{{Suzuki} {et~al.}(2003){Suzuki}, {Tytler}, {Kirkman}, {O'Meara}, \&
  {Lubin}}]{Suzuki-03}
{Suzuki}, N., {Tytler}, D., {Kirkman}, D., {O'Meara}, J.~M., \& {Lubin}, D.
  2003, \pasp, 115, 1050

\bibitem[{{{\v{S}}koda} \& {Hensberge}(2003)}]{Skoda-03}
{{\v{S}}koda}, P. \& {Hensberge}, H. 2003, in Astronomical Society of the
  Pacific Conference Series, Vol. 295, Astronomical Data Analysis Software and
  Systems XII, ed. H.~E. {Payne}, R.~I. {Jedrzejewski}, \& R.~N. {Hook}, 415

\bibitem[{{Yanny} {et~al.}(2009){Yanny}, {Rockosi}, {Newberg}, {Knapp},
  {Adelman-McCarthy}, {Alcorn}, {Allam}, {Allende Prieto}, {An}, {Anderson},
  {Anderson}, {Bailer-Jones}, {Bastian}, {Beers}, {Bell}, {Belokurov},
  {Bizyaev}, {Blythe}, {Bochanski}, {Boroski}, {Brinchmann}, {Brinkmann},
  {Brewington}, {Carey}, {Cudworth}, {Evans}, {Evans}, {Gates}, {G{\"a}nsicke},
  {Gillespie}, {Gilmore}, {Nebot Gomez-Moran}, {Grebel}, {Greenwell}, {Gunn},
  {Jordan}, {Jordan}, {Harding}, {Harris}, {Hendry}, {Holder}, {Ivans},
  {Ivezi{\v{c}}}, {Jester}, {Johnson}, {Kent}, {Kleinman}, {Kniazev},
  {Krzesinski}, {Kron}, {Kuropatkin}, {Lebedeva}, {Lee}, {French Leger},
  {L{\'e}pine}, {Levine}, {Lin}, {Long}, {Loomis}, {Lupton}, {Malanushenko},
  {Malanushenko}, {Margon}, {Martinez-Delgado}, {McGehee}, {Monet}, {Morrison},
  {Munn}, {Neilsen}, {Nitta}, {Norris}, {Oravetz}, {Owen}, {Padmanabhan},
  {Pan}, {Peterson}, {Pier}, {Platson}, {Re Fiorentin}, {Richards}, {Rix},
  {Schlegel}, {Schneider}, {Schreiber}, {Schwope}, {Sibley}, {Simmons},
  {Snedden}, {Allyn Smith}, {Stark}, {Stauffer}, {Steinmetz}, {Stoughton},
  {SubbaRao}, {Szalay}, {Szkody}, {Thakar}, {Sivarani}, {Tucker}, {Uomoto},
  {Vanden Berk}, {Vidrih}, {Wadadekar}, {Watters}, {Wilhelm}, {Wyse}, {Yarger},
  \& {Zucker}}]{Yanny-09}
{Yanny}, B., {Rockosi}, C., {Newberg}, H.~J., {et~al.} 2009, \aj, 137, 4377

\end{thebibliography}


\begin{appendix}

\section{Additional tables}

\begin{table*}[]
\caption{\label{tab:all_parameters} Main properties of the Gaia benchmark stars with ESPRESSO spectra. The stellar parameters are from \citet{Jofre-15}.}
\centering
\small
\begin{tabular}{llllll}
\hline\hline
           star &  S/N(ESPRESSO) & \teff &   \logg &  [Fe/H] & v$\sin i$ \\
\hline
         18 Sco &           570 &     5810$\pm$80 &     4.44$\pm$0.03 &   0.01$\pm$0.03 &      2.2$\pm$1.2 \\
       Arcturus &           570 &     4286$\pm$35 &     1.64$\pm$0.09 &  -0.53$\pm$0.08 &      3.8$\pm$1.0 \\
 $\alpha$ Cen A &           620 &     5792$\pm$16 &     4.31$\pm$0.01 &   0.24$\pm$0.08 &      1.9$\pm$0.6 \\
 $\alpha$ Cen B &           680 &     5231$\pm$20 &     4.53$\pm$0.03 &    0.22$\pm$0.1 &      1.0$\pm$0.6 \\
   $\alpha$ Cet &           360 &     3796$\pm$65 &     0.68$\pm$0.23 &  -0.45$\pm$0.47 &      3.0$\pm$2.0 \\
        Procyon &           380 &     6554$\pm$84 &      4.0$\pm$0.02 &  -0.04$\pm$0.08 &      2.8$\pm$0.6 \\
      Aldebaran &           670 &     3927$\pm$40 &     1.11$\pm$0.19 &  -0.37$\pm$0.17 &      5.0$\pm$1.0 \\
    $\beta$ Ara &           490 &     4197$\pm$50 &     1.05$\pm$0.15 &  -0.05$\pm$0.39 &      5.4$\pm$1.0 \\
    $\beta$ Gem &           520 &     4858$\pm$60 &      2.9$\pm$0.08 &   0.12$\pm$0.16 &      2.0$\pm$1.0 \\
    $\beta$ Hyi &           520 &     5873$\pm$45 &     3.98$\pm$0.02 &  -0.07$\pm$0.06 &      3.3$\pm$0.3 \\
    $\beta$ Vir &           690 &     6083$\pm$41 &      4.1$\pm$0.02 &   0.21$\pm$0.07 &      2.0$\pm$0.6 \\
   $\delta$ Eri &           800 &     4954$\pm$30 &     3.76$\pm$0.02 &   0.06$\pm$0.05 &      0.7$\pm$0.6 \\
 $\epsilon$ Eri &          1510 &     5076$\pm$30 &     4.61$\pm$0.03 &   -0.1$\pm$0.06 &      2.4$\pm$0.2 \\
 $\epsilon$ For &           540 &     5123$\pm$78 &     3.52$\pm$0.08 &   -0.62$\pm$0.1 &      4.2$\pm$1.0 \\
 $\epsilon$ Vir &           570 &     4983$\pm$61 &     2.77$\pm$0.02 &   0.13$\pm$0.16 &      2.0$\pm$1.0 \\
     $\eta$ Boo &           700 &     6099$\pm$28 &     3.79$\pm$0.02 &    0.3$\pm$0.08 &     12.7$\pm$1.4 \\
   $\gamma$ Sge &           590 &     3807$\pm$49 &     1.05$\pm$0.32 &  -0.16$\pm$0.39 &      6.0$\pm$1.0 \\
      HD 107328 &           600 &     4496$\pm$59 &     2.09$\pm$0.13 &  -0.34$\pm$0.16 &      1.9$\pm$1.2 \\
       HD122563 &           840 &     4587$\pm$60 &     1.61$\pm$0.07 &  -2.74$\pm$0.22 &      5.0$\pm$2.0 \\
       HD140283 &           440 &    5522$\pm$105 &     3.58$\pm$0.11 &   -2.43$\pm$0.1 &      5.0$\pm$2.0 \\
       HD220009 &           400 &     4217$\pm$59 &     1.43$\pm$0.12 &  -0.75$\pm$0.13 &      1.0$\pm$1.0 \\
        HD22879 &           590 &     5868$\pm$89 &     4.27$\pm$0.04 &  -0.88$\pm$0.05 &      4.4$\pm$1.0 \\
        HD49933 &           610 &     6635$\pm$91 &      4.2$\pm$0.03 &  -0.46$\pm$0.08 &     10.0$\pm$0.5 \\
        HD84937 &           560 &     6356$\pm$97 &     4.06$\pm$0.04 &  -2.09$\pm$0.08 &      5.2$\pm$2.0 \\
      $\xi$ Hya &           550 &     5044$\pm$40 &     2.87$\pm$0.02 &    0.14$\pm$0.2 &      2.4$\pm$0.6 \\
      $\mu$ Ara &           480 &     5902$\pm$66 &      4.3$\pm$0.03 &   0.33$\pm$0.13 &      2.2$\pm$0.8 \\
      $\mu$ Leo &           460 &     4474$\pm$60 &     2.51$\pm$0.11 &   0.26$\pm$0.15 &      5.1$\pm$1.0 \\
     $\psi$ Phe &           330 &     3472$\pm$92 &     0.51$\pm$0.18 &  -1.23$\pm$0.39 &      3.0$\pm$1.0 \\
            Sun &           530 &      5771$\pm$1 &  4.438$\pm$0.0002 &   0.02$\pm$0.05 &      1.6$\pm$0.3 \\
     $\tau$ Cet &           570 &     5414$\pm$21 &     4.49$\pm$0.02 &   -0.5$\pm$0.03 &      0.4$\pm$0.4 \\
\hline
\end{tabular}
\end{table*}

\begin{table*}[t!]
\caption{Stellar parameters and chemical abundances of the Sun derived from the  ESPRESSO combined and the individual 1D spectra (from Spec 1 to Spec 5 in the table). \label{tab:sun_parameters_abundances} }
\centering
\footnotesize
\begin{tabular}{lcccccc}
\hline\hline
Parameter & ESPRESSO & Spec 1 & Spec 2 & Spec 3 & Spec 4 & Spec 5  \\
\hline
\teff  &  5768$\pm$12  &  5782$\pm$11  &  5781$\pm$11  &  5783$\pm$10  &  5780$\pm$11  &  5760$\pm$10 \\
\logg  &  4.410$\pm$0.027  &  4.445$\pm$0.033  &  4.415$\pm$0.027  &  4.435$\pm$0.020  &  4.423$\pm$0.020  &  4.423$\pm$0.020\\
\vtur  &  0.968$\pm$0.017  &  1.010$\pm$0.018  &  1.024$\pm$0.018  &  0.996$\pm$0.015  &  0.975$\pm$0.017  &  0.981$\pm$0.014\\
{[}Fe/H{]}  &  0.004$\pm$0.010  &  0.003$\pm$0.009  &  0.003$\pm$0.009  &  0.007$\pm$0.008  &  0.008$\pm$0.009  &  -0.005$\pm$0.008\\
{[}NaI/H{]}  &  0.053$\pm$0.039  &  0.066$\pm$0.047  &  0.069$\pm$0.059  &  0.076$\pm$0.042  &  0.082$\pm$0.038  &  0.063$\pm$0.044 \\ 
{[}MgI/H{]}  &  0.025$\pm$0.016  &  0.026$\pm$0.067  &  0.037$\pm$0.075  &  0.026$\pm$0.053  &  0.033$\pm$0.047  &  0.024$\pm$0.048 \\ 
{[}AlI/H{]}  &  0.014$\pm$0.019  &  0.028$\pm$0.074  &  0.020$\pm$0.095  &  0.011$\pm$0.065  &  0.020$\pm$0.056  &  -0.004$\pm$0.068 \\ 
{[}SiI/H{]}  &  0.006$\pm$0.019  &  0.012$\pm$0.008  &  0.016$\pm$0.019  &  0.020$\pm$0.010  &  0.028$\pm$0.013  &  0.021$\pm$0.013 \\ 
{[}CaI/H{]}  &  0.027$\pm$0.043  &  -0.002$\pm$0.041  &  0.017$\pm$0.039  &  0.027$\pm$0.035  &  0.030$\pm$0.039  &  0.015$\pm$0.037 \\ 
{[}TiI/H{]}  &  0.003$\pm$0.023  &  0.012$\pm$0.017  &  0.015$\pm$0.024  &  0.022$\pm$0.016  &  0.017$\pm$0.021  &  0.004$\pm$0.022 \\ 
{[}TiII/H{]}  &  -0.009$\pm$0.014  &  -0.006$\pm$0.018  &  -0.009$\pm$0.015  &  -0.006$\pm$0.014  &  -0.007$\pm$0.010  &  -0.003$\pm$0.014 \\ 
{[}CrI/H{]}  &  0.038$\pm$0.025  &  0.039$\pm$0.026  &  0.037$\pm$0.022  &  0.038$\pm$0.023  &  0.028$\pm$0.024  &  0.024$\pm$0.022 \\ 
{[}NiI/H{]}  &  0.004$\pm$0.013  &  0.013$\pm$0.014  &  0.015$\pm$0.012  &  0.012$\pm$0.014  &  0.012$\pm$0.013  &  0.000$\pm$0.013 \\ 
\hline
\end{tabular}
\end{table*}

\section{Two slices of ESPRESSO S2D spectra}                                    \label{sec:slices}

\begin{figure*}
\begin{center}
\begin{tabular}{cc}
\includegraphics[angle=0,width=0.45\linewidth]{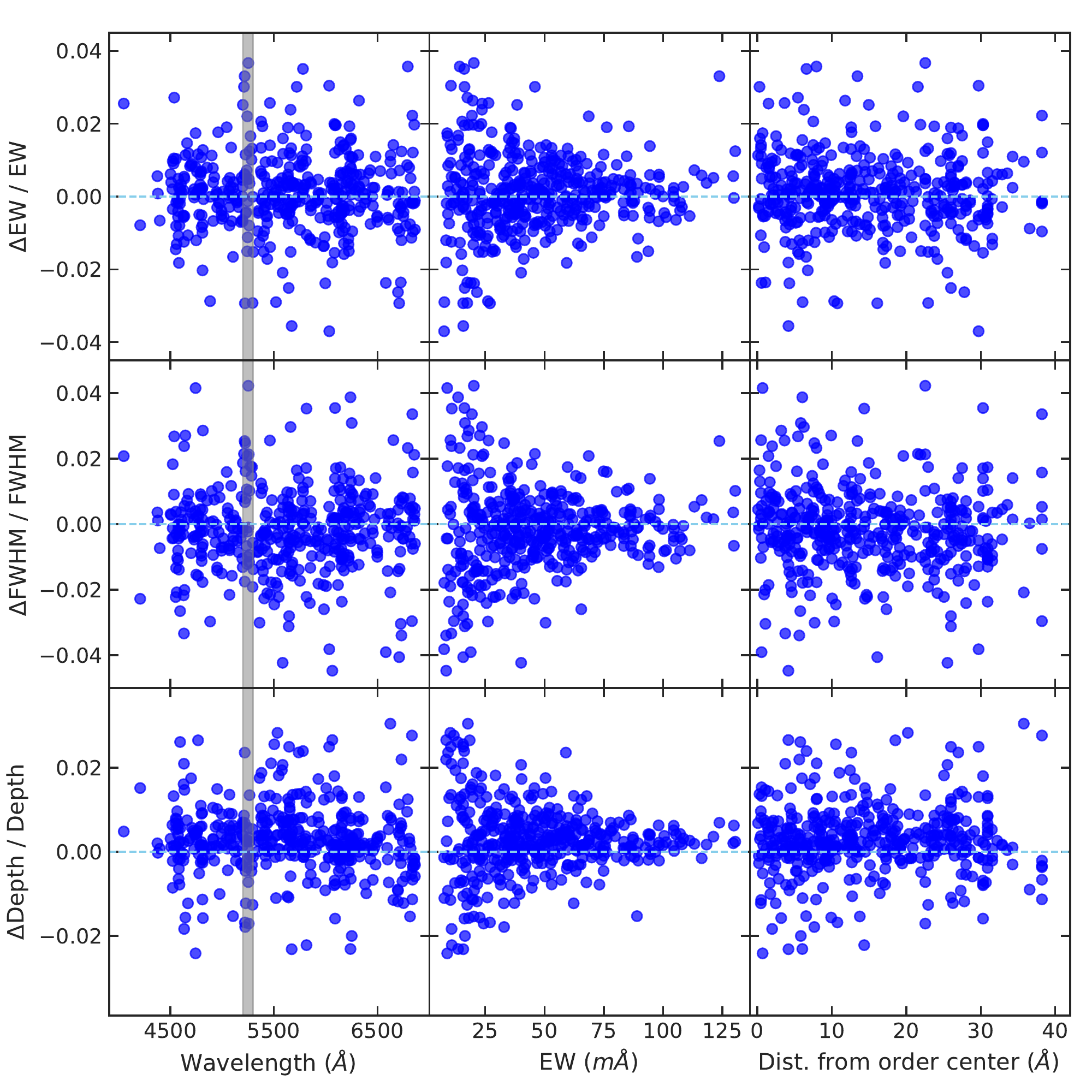} &
\includegraphics[angle=0,width=0.45\linewidth]{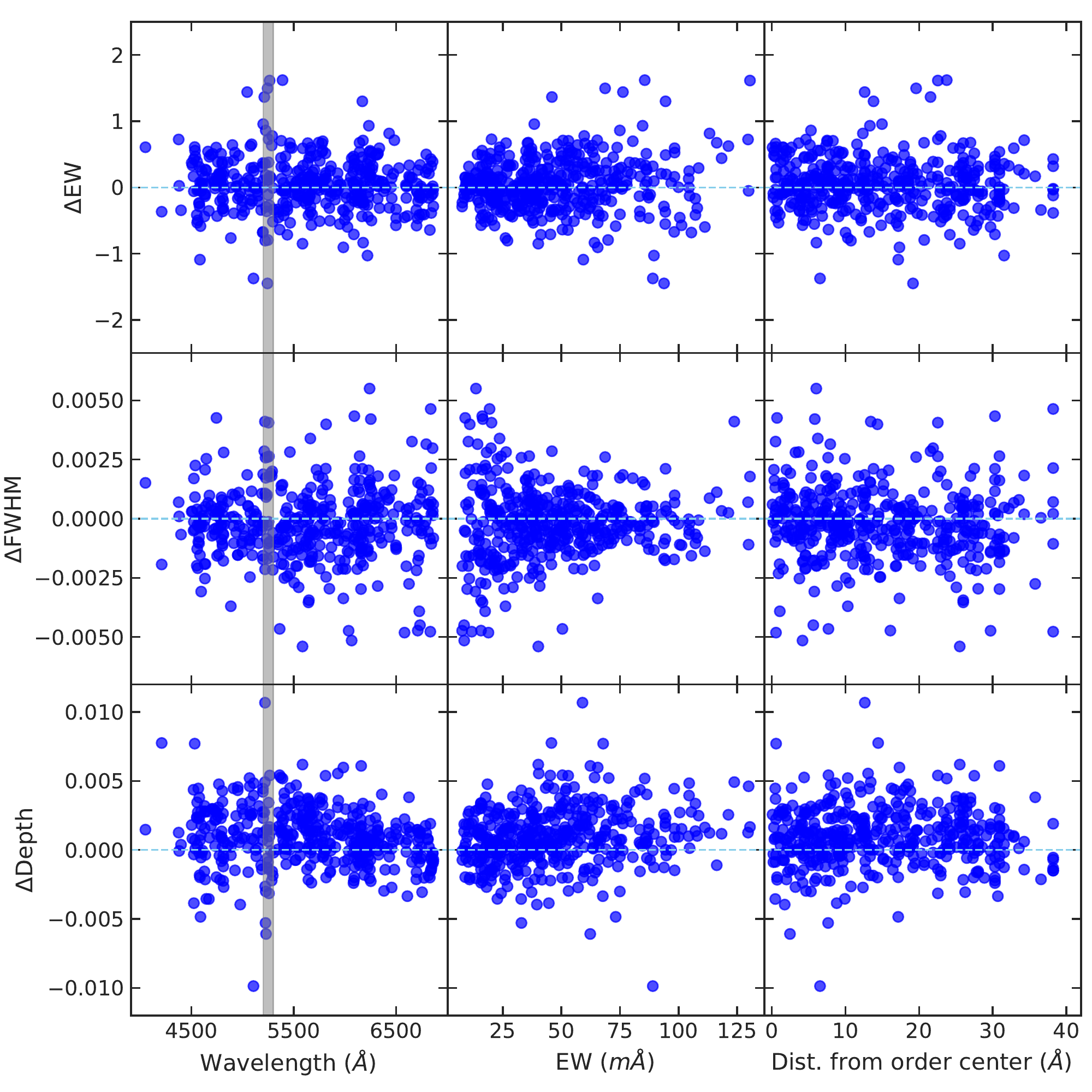}
\end{tabular}\end{center}
\vspace{-0.5cm}
\caption{Relative (left panel) and absolute (right panel) differences of spectral line parameters measured in two slices of the S2D spectra as a function of wavelength, EW, and the distance from the center of the spectral order. The boundary wavelength region (at  $\sim$5250\,\AA{}) of the blue- and red-arm detectors is indicated by the rectangle. Line parameters were measured with ARES with a S/N that varied as a function of the position of the spectral line in the echelle order.}
\label{fig-sun_slices}
\end{figure*}

As mentioned in Sect.~\ref{sec:espresso},  two simultaneously observed 2D spectra are obtained with the pupil slicer. In this section we compare the spectral line parameters as measured from the two slices of the S2D spectra. The line measurements were performed as in the previous section, considering a variable S/N as a function of the position of the spectral lines. The results of our measurements are shown in Fig.~\ref{fig-sun_slices}. As in the plots of the previous section, the average values obtained for the five solar spectra are shown. The outlier lines (at a 3$\sigma$ level) were removed. For the spectral lines observed in different orders, the mean values are considered.

The left panel of Fig.~\ref{fig-sun_slices} shows that the line parameter measurements from the two slices are consistent within a few percent. No trend can be seen in the figure, except for the fact that the relative difference in EW measurements increases for weak lines. This is expected because the absolute differences in EWs are nearly constant; they are $\Delta$EW = 0.05$\pm$0.44 m\AA{}. 

In the previously discussed plot we did not separate the spectral lines observed in only single orders from the lines observed in multiple orders because we found no difference between the two groups: 0.1$\pm$1.0\% vs 0.1$\pm$1.1\% for EW, 0.1$\pm$1.2\% vs 0.1$\pm$1.2\% for FWHM, and 0.2$\pm$0.8\% vs 0.2$\pm$0.9\% for line depth.
While we found practically no difference in EWs of the lines as measured from the two slices and relatively small scatter of 0.44 m\AA{} for $\Delta$EW, we decided to perform additional tests to understand which fraction of this scatter comes from our method of measurements of the lines and which fraction is due to the limited S/N of the spectra (slices). We compared the EWs as measured from the two slices for each of the five solar spectra. The average difference was nearly zero for all the cases, and the scatter ranged from 0.74 m\AA{} (for the poorest S/N spectrum) to 0.98 m\AA{} (for the highest quality spectrum). Following \citet{Cayrel-88}, we calculated the average statistical photonic error for the line EWs. This error estimate is the theoretical lower limit because it does not take various sources of error into consideration, such as continuum placement. The Cayrel formulae suggest an average error of 0.41 m\AA{} (to be compared with 0.74 m\AA{}) for the spectra with the highest S/N and an error of 0.63 m\AA{} (to be compared with 0.98 m\AA{}) for the lowest S/N solar spectrum. This simple comparison shows that ARES measurements of the EWs of the spectral lines on average are commensurate with the statistical photonic errors in EWs expected for the S/N of the spectra.

\section{Spectrum-to-spectrum scatter}                                  \label{sec:1d_1d}

\begin{figure*}
\begin{center}
\begin{tabular}{cc}
\includegraphics[angle=0,width=0.45\linewidth]{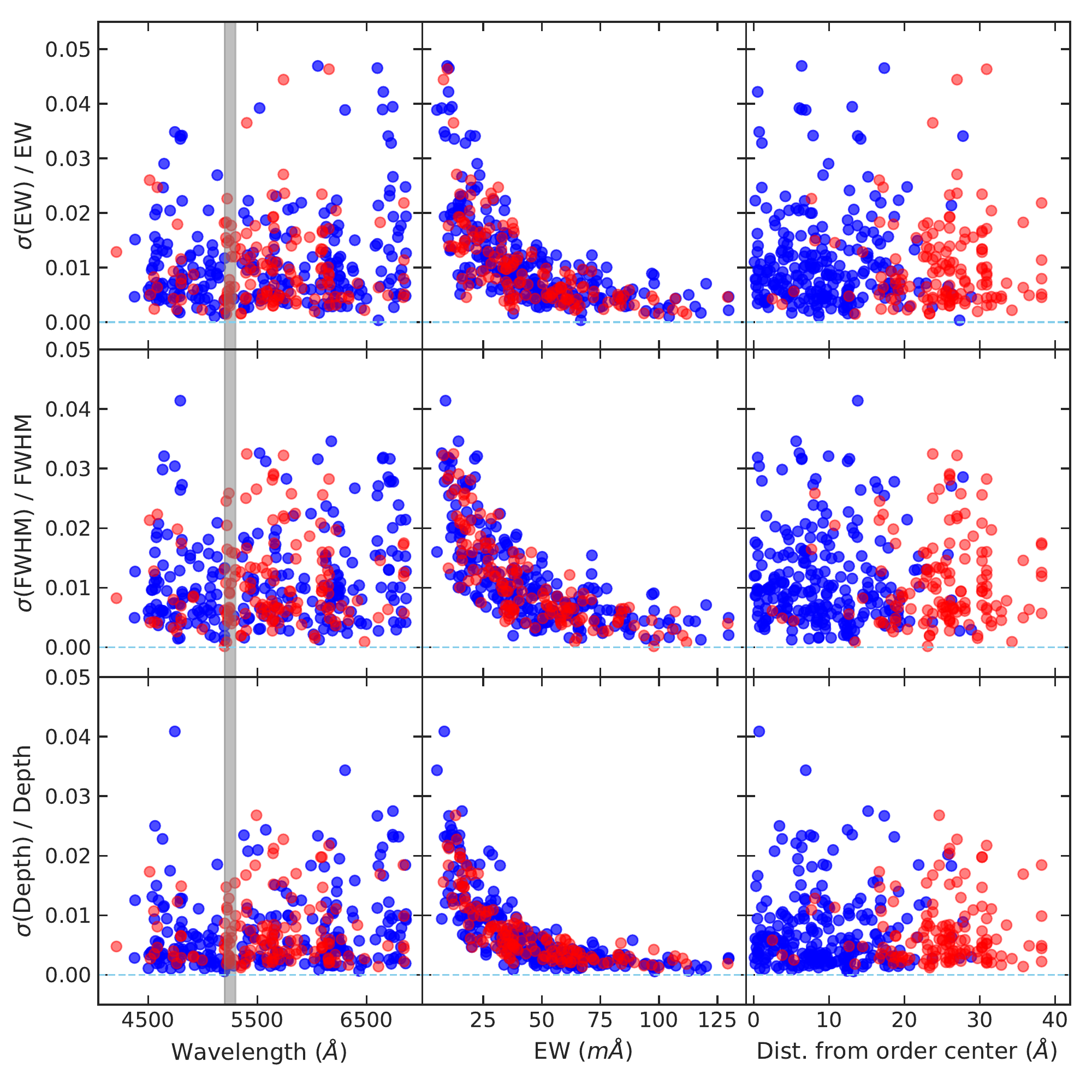} &
\includegraphics[angle=0,width=0.45\linewidth]{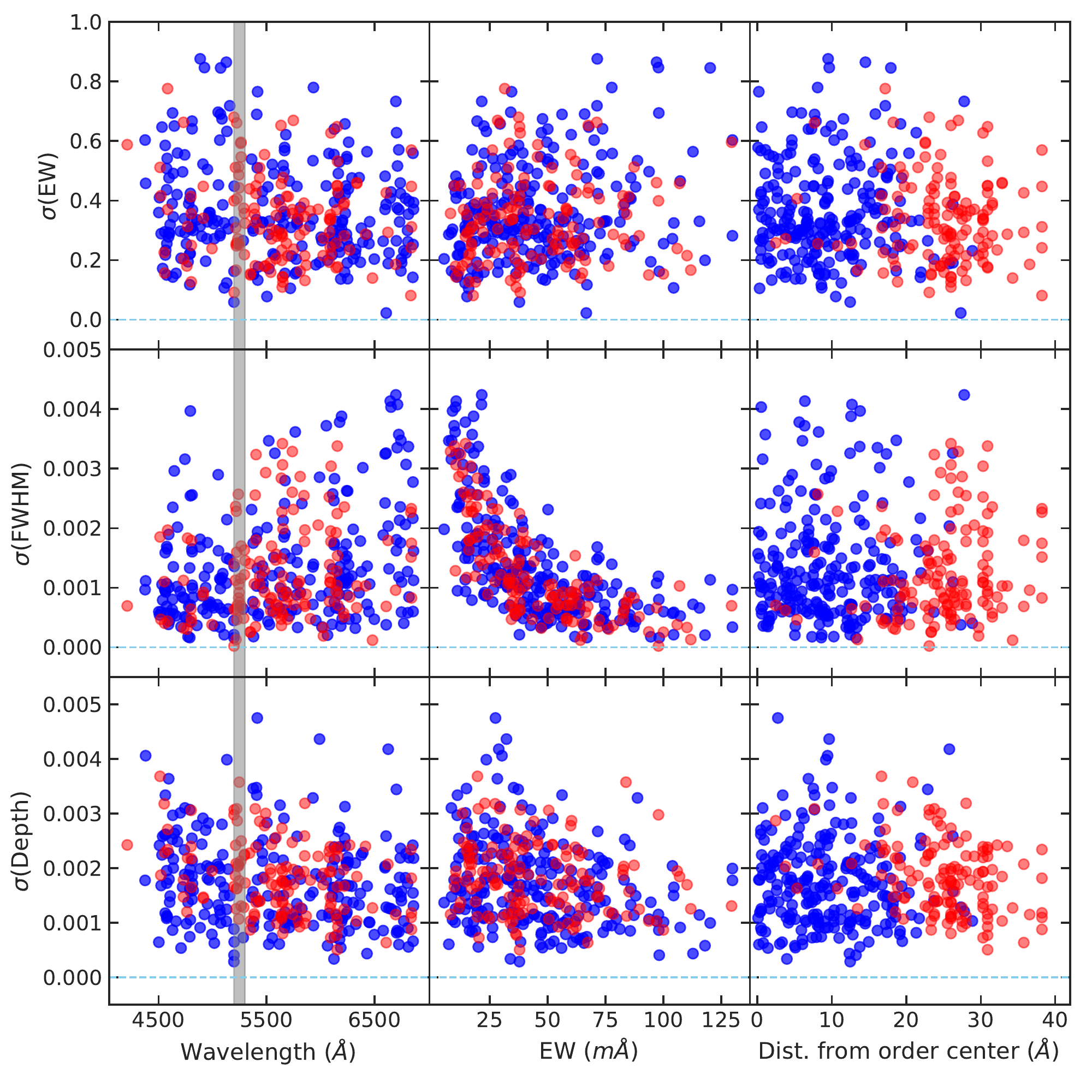}
\end{tabular}\end{center}
\vspace{-0.5cm}
\caption{Relative (left panel) and absolute (right panel) spectrum-to-spectrum scatter of spectral line parameters measured from five 1D solar spectra as a function of wavelength, EW, and the distance from the center of the spectral order. The red and blue symbols correspond to the lines that in the 2D spectrum are observed in multiple and single orders, respectively. The boundary wavelength region (at  $\sim$5250\,\AA{}) of the blue- and red-arm detectors is indicated by the rectangle.}
\label{fig-sun_s1d_s1d}
\end{figure*}

The next test we performed for the five UHR solar spectra was to compare the line parameters as measured from these five S1D spectra. As in the previous sections, we discarded 3$\sigma$  outlier lines, that is, lines for which the spectrum-to-spectrum scatter for a given line parameter is at least three times larger than the average of spectrum-to-spectrum scatter for all the lines. The results of our measurements are shown in Fig.~\ref{fig-sun_s1d_s1d}. The plot shows no difference between lines located (in 2D spectra) in single and multiple orders.

In general, the line measurements are highly consistent. The average relative scatter for EW is 1.1$\pm$0.8\% and increases only to about 4\% for the weakest lines with EW $<$ 10  m\AA{}. The absolute scatter for the EW is 0.35$\pm$0.15 m\AA,{} which is very close to the average error of 0.3 m\AA{} in EW estimated by the \citet{Cayrel-88} formulae for our spectra.

The next logical step in our study is to estimate the effect of the spectrum-to-spectrum, non-astrophysical variations of the line parameters on the stellar parameter determinations. We determined the atmospheric parameters of the Sun from the five individual S1D spectra following \citet{Sousa-15a}. In brief, we used the EWs of the FeI and FeII lines and searched for the ionization and excitation equilibrium in our spectral analysis. We used the grid of Kurucz model atmospheres \citep{Kurucz-93} and the 2014 version of the radiative transfer code MOOG \citep{Sneden-73}. We refer  to \citet{Sousa-14} for more details about the method. 

The results are presented in Table~\ref{tab:sun_parameters_abundances}. The parameter determinations agree excellently well in general. In particular, the observed spectrum-to-spectrum scatter for \teff \ is $\sim$9 K, while the average error is $\sim$11 K, for \logg \ the scatter is $\sim$0.01 dex with an average error of $\sim$0.02 dex, for \vtur \ the scatter is $\sim$0.02 km/s with an average error of $\sim$0.02 km/s, and finally, for the [Fe/H] the observed scatter is $\sim$0.005 dex, while the average error is $\sim$0.009 dex. This result also suggest that the procedure we used to determine the uncertainties (internal precision) is consistent and robust.

\begin{figure}
\begin{center}
\begin{tabular}{c}
\includegraphics[angle=0,width=0.85\linewidth]{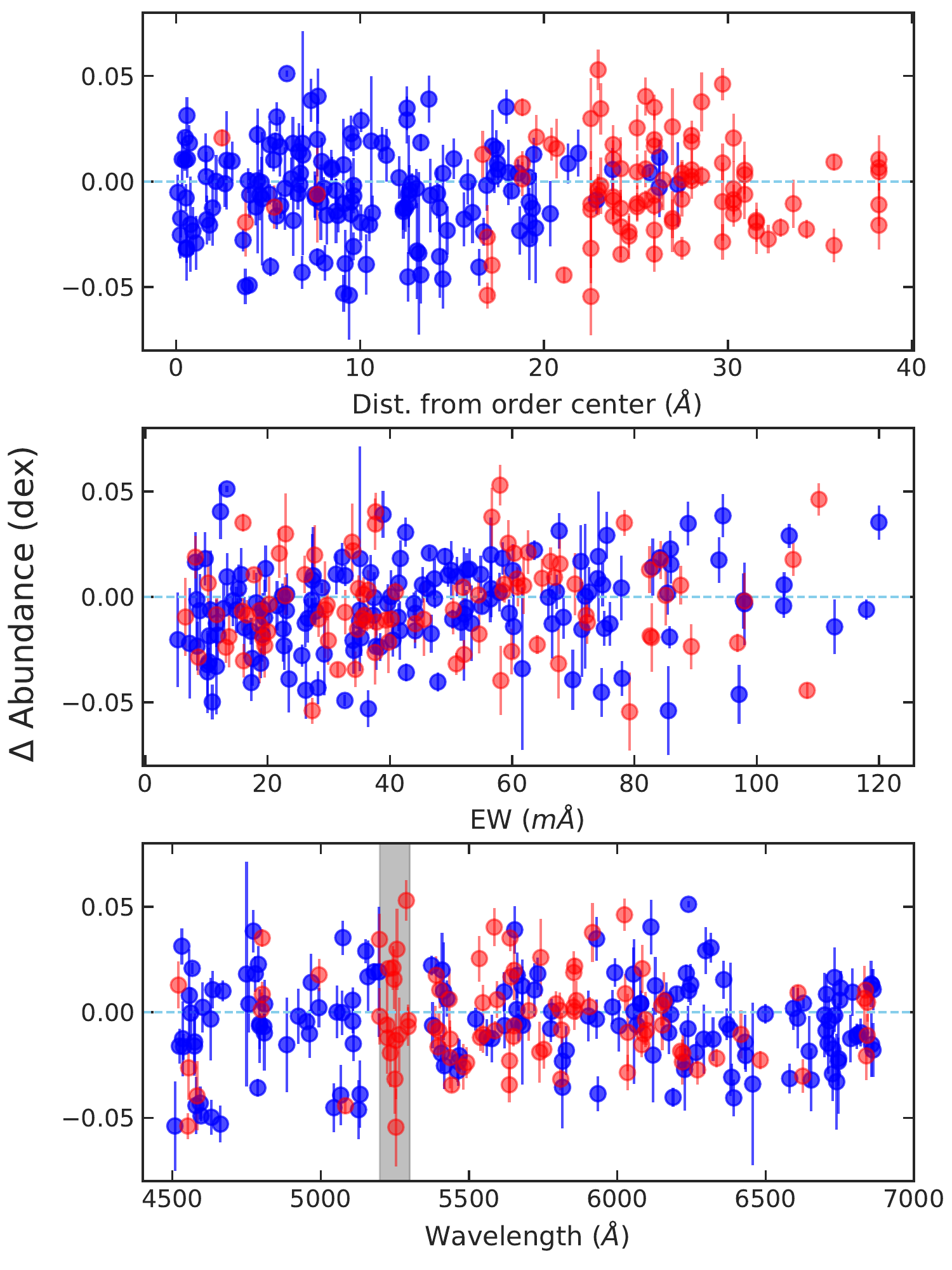}
\end{tabular}\end{center}
\vspace{-0.5cm}
\caption{Deviation of individual (derived from individual spectral lines) iron abundances from the mean abundance as a function of wavelength, EW, and the distance from the center of the spectral order. The symbols show the average abundances for the five solar spectra and the error-bars represent the standard deviation of the five determinations. Outliers at 3-$\sigma$ level are excluded. The red and blue symbols correspond to the lines that in the 2D spectrum are observed in multiple orders and in single orders, respectively. The boundary wavelength region (at  $\sim$5250\,\AA{}) of the blue- and red-arm detectors is indicated by the rectangle. EW measurements are performed on the S1D spectra.}
\label{fig-sun_rel_s1d_s1d_fe}
\end{figure}

Because our iron line-list is composed of a large number of  single and  multiple lines, it gives us a good opportunity to test whether the two sets of lines deliver statistically discrepant abundances. In Fig.~\ref{fig-sun_rel_s1d_s1d_fe} we show how the abundances of individual lines deviate from the mean iron abundance as a function of their EW, wavelength, and distance from the center of the spectral order. For each line, the average abundance derived from the five solar S1D spectra is considered. The figure does not show any particular trend for lines observed in single or multiple orders. The average difference between the abundances derived from these two sets of lines is $\sim$0.001$\pm$0.015 dex.  The scatter around the mean is slightly smaller for the  multiple lines ($\sim$0.010 dex) than for the  single lines ($\sim$0.012 dex).

\begin{figure}
\begin{center}
\begin{tabular}{c}
\includegraphics[angle=0,width=0.85\linewidth]{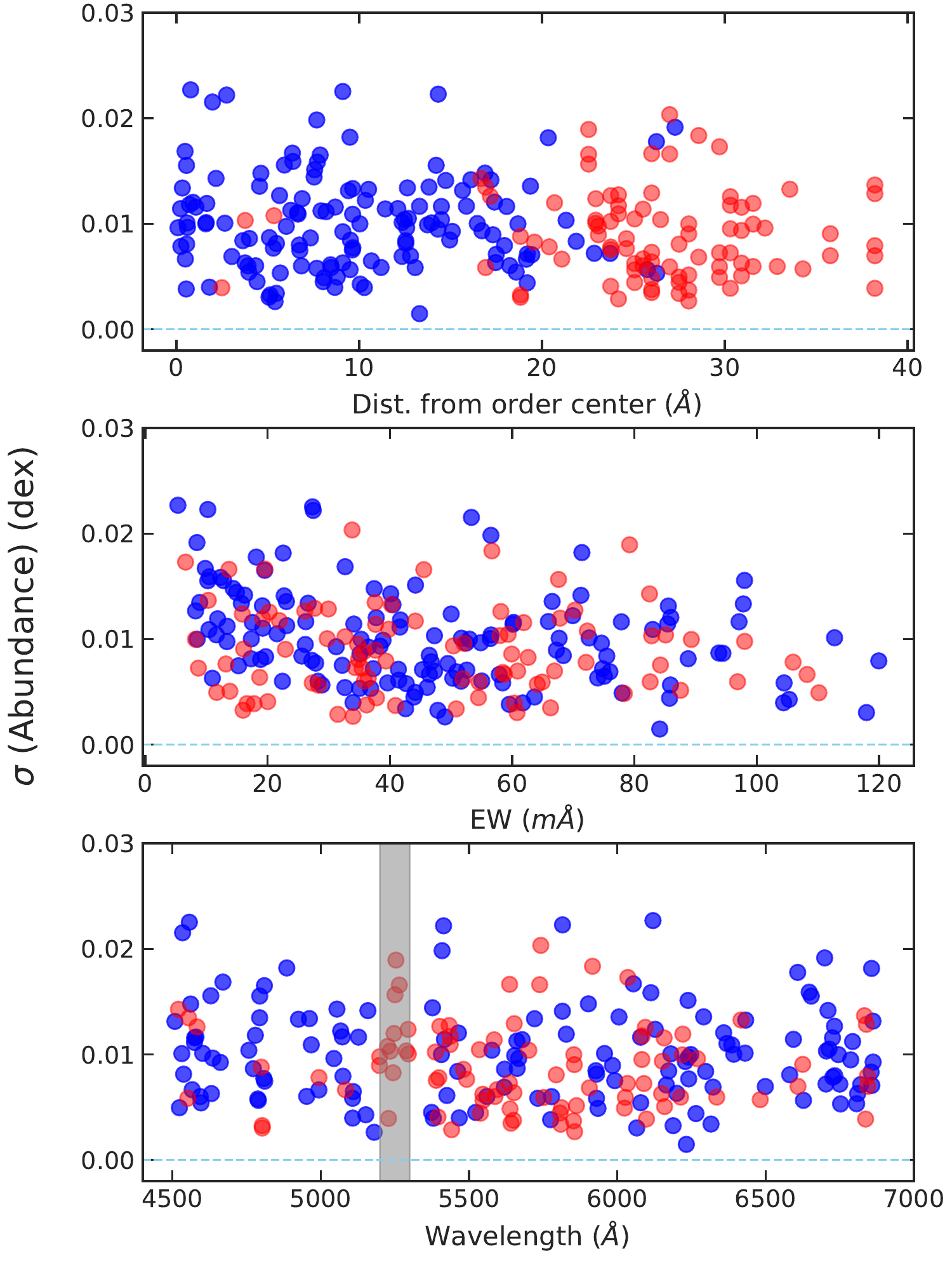}
\end{tabular}\end{center}
\vspace{-0.5cm}
\caption{Spectrum-to-spectrum abundance scatter of individual iron lines as a function of wavelength, EW, and the distance from the center of the spectral order. Outliers at 3$\sigma$ level are excluded. The red and blue symbols correspond to the lines that in the 2D spectrum are observed in multiple and single orders, respectively. The boundary wavelength region (at  $\sim$5250\,\AA{}) of the blue- and red-arm detectors is indicated by the rectangle. EW measurements are performed on the S1D spectra.}
\label{fig-sun_rel_s1d_s1d_fe_abund}
\end{figure}

In Fig.~\ref{fig-sun_rel_s1d_s1d_fe_abund} we show the spectrum-to-spectrum abundance scatter for each iron line. The mean scatter for single and multiple lines is very similar: 0.010$\pm$0.004 dex and 0.009$\pm$0.004 dex, respectively. The mean scatter for all the lines is 0.009$\pm$0.004 dex, which can be considered as the translation of the average spectrum-to-spectrum  EW scatter of 0.35$\pm$0.16 m\AA{} into iron abundances.

\end{appendix}

\end{document}